\documentclass[aps,prd,preprint,tightenlines,groupedaddress,nofootinbib]{revtex4}
\usepackage{amssymb,latexsym}
\usepackage{amsmath,amsbsy}
\usepackage{epsfig,bm}
\usepackage{graphicx,comment}
\DeclareMathOperator{\tr}{tr}

\begin{document}
\def\a{{\alpha}}
\def\b{{\beta}}
\def\d{{\delta}}
\def\D{{\Delta}}
\def\e{{\epsilon}}
\def\g{{\gamma}}
\def\G{{\Gamma}}
\def\k{{\kappa}}
\def\l{{\lambda}}
\def\L{{\Lambda}}
\def\m{{\mu}}
\def\n{{\nu}}
\def\o{{\omega}}
\def\O{{\Omega}}
\def\S{{\Sigma}}
\def\s{{\sigma}}
\def\th{{\theta}}

\def\ol#1{{\overline{#1}}}

\def\Dslash{D\hskip-0.65em /}

\def\CPT{{$\chi$PT}}
\def\QCPT{{Q$\chi$PT}}
\def\PQCPT{{PQ$\chi$PT}}
\def\tr{\text{tr}}
\def\str{\text{str}}
\def\diag{\text{diag}}
\def\order{{\mathcal O}}

\def\cC{{\mathcal C}}
\def\cB{{\mathcal B}}
\def\cT{{\mathcal T}}
\def\cQ{{\mathcal Q}}
\def\cL{{\mathcal L}}
\def\cO{{\mathcal O}}
\def\cA{{\mathcal A}}
\def\cQ{{\mathcal Q}}
\def\cR{{\mathcal R}}
\def\cH{{\mathcal H}}

\def\eqref#1{{(\ref{#1})}}

\preprint{NT@UW 03-030}
\title{Hadronic Electromagnetic Properties at Finite Lattice Spacing}
\author{Daniel Arndt}
\email[]{arndt@phys.washington.edu}
\author{Brian C.~Tiburzi}
\email[]{bctiburz@phys.washington.edu}
\affiliation{Department of Physics,  	Box 351560,\\
	University of Washington,     
	Seattle, WA 98195-1560, USA}
\date{\today}

\begin{abstract}
Electromagnetic properties of the octet mesons as well as the octet and decuplet baryons 
are augmented 
in quenched and partially quenched chiral perturbation theory
to include 
$\cO(a)$ corrections due to lattice discretization. 
We present the results for the $SU(3)$ flavor 
group in the isospin limit as well as the results for $SU(2)$ flavor 
with non-degenerate quarks. These corrections will be useful for 
extrapolation of lattice calculations using Wilson valence and sea quarks, 
as well as calculations using Wilson sea quarks and Ginsparg-Wilson valence quarks. 
\end{abstract}

\maketitle

\section{Introduction}
Lattice gauge theory can provide first principle calculations in the strongly 
interacting regime of QCD, where quarks and gluons are bound into color-neutral hadronic states. 
These calculations, however, are severely limited by the available computing power, 
necessitating the use of light quark masses $m_q$ that are much larger than those in reality. 
Hence, one needs to extrapolate from the quark masses used on the lattice to those of nature.

A model independent way to do this extrapolation is to study QCD at hadronic scales 
through its low-energy effective theory, 
chiral perturbation theory (\CPT).
Since \CPT\ provides a systematic expansion in terms of $m_q/\L_\chi$, where $\L_\chi$ is 
the chiral symmetry breaking scale, one can, in principle, understand how QCD observables 
behave as functions of the quark mass. 
In order to address the quenched and partially quenched approximations employed by
lattice calculations, \CPT\ has been extended to quenched chiral 
perturbation theory (\QCPT)~%
\cite{Morel:1987xk,Sharpe:1992ft,Bernard:1992mk,Bernard:1992ep,Golterman:1994mk,Labrenz:1996jy,Sharpe:1996qp}
and partially quenched chiral perturbation theory (\PQCPT)~%
\cite{Bernard:1994sv,Sharpe:1997by,Golterman:1998st,Sharpe:1999kj,Sharpe:2000bn,Sharpe:2000bc,Sharpe:2001fh,Shoresh:2001ha,Sharpe:2003vy}.

Recently, we considered the electromagnetic properties of the octet mesons and both the 
octet and decuplet baryons in \QCPT\ and \PQCPT~%
\cite{Arndt:2003ww,Arndt:2003we,Arndt:2003vd}.
Owing in part to the charge neutrality of singlet fields, the quenched results are not 
more singular in the chiral limit than their unquenched counterparts. We showed, however,
that despite this similarity, the quenched results contain singlet contributions that have 
no analog in \CPT. Moreover, quenching closed quark loops alters the
contribution from chiral logs. For the decuplet baryon form factors, for example, quenching completely
removes these chiral logs. Many others have also observed that the behavior of meson loops near the chiral limit
is misrepresented in \QCPT, see for example~%
\cite{Booth:1995hx,Kim:1998bz,Savage:2001dy,Arndt:2002ed,Dong:2003im}. 
On the other hand, \PQCPT\ results are devoid of such complications 
and allow for a smooth limit to QCD. 

Not only are lattice calculations limited to unphysically large quark masses, they are also severely restricted by
two further parameters: the size $L$ of the lattice, that is not considerably larger than the system under investigation; 
and the lattice spacing $a$, that is not considerably smaller than the relevant hadronic distance scale.
To address the issue of finite lattice spacing,
\CPT\ has been extended (following the earlier work of~\cite{Sharpe:1998xm,Lee:1999zx})
in the meson sector
to $\cO(a)$ for the Wilson action~\cite{Rupak:2002sm} and for mixed actions~\cite{Bar:2002nr}.
Corrections at $\cO(a^2)$ have also been pursued~\cite{Bar:2003mh,Aoki:2003yv}.
Corrections to baryon 
observables have only recently been investigated~\cite{Beane:2003xv}. 
To consider finite lattice spacing corrections, one must formulate the underlying lattice theory and match
the new operators that appear onto those in the chiral effective theory. This can be done by utilizing a dual 
expansion in quark mass and lattice spacing.
Following~\cite{Bar:2003mh,Beane:2003xv}, we assume a hierarchy of energy scales
\begin{equation}
m_q \ll \L_\chi \ll \frac{1}{a}
\end{equation}
and ignore finite volume effects.
The small dimensionless expansion parameters are
\begin{equation} \label{eqn:pc}
\e^2 \sim 
\begin{cases}
 m_q/\L_\chi, \\
 a \, \L_\chi, \\ 
 p^2/\L_\chi^2
\end{cases}
\end{equation}
where $p$ is an external momentum. Thus we have a systematic way to calculate $\cO(a)$ corrections
in \CPT\ for the observables of interest. 

In this work we investigate the $\cO(a)$ corrections to the electromagnetic properties 
of the meson and baryon octets, the baryon decuplet, and the decuplet to octet electromagnetic
transitions
in \QCPT\ and \PQCPT.
We work up to next-to-leading order in the chiral expansion
and to leading order in the heavy baryon expansion.
The paper is structured as follows. First, in Section~\ref{sec:PQCPT}, we 
review \PQCPT\ at finite lattice spacing with mixed actions.
Since the setup for \QCPT\ parallels that of \PQCPT,
we will only highlight differences where appropriate. 
Next in Section~\ref{sec:mesons} we calculate finite lattice spacing corrections to 
the charge radii of the octet mesons to $\cO(\e^2)$. 
This is followed by the calculation of such corrections to:
the charge radii and magnetic moments of the octet baryons; 
the charge radii, magnetic moments, and electric quadrupole moments of the decuplet
baryons;
and the decuplet to octet electromagnetic transition moments (Sections~\ref{sec:octet}--\ref{sec:trans}). 
Corresponding results for the above 
electromagnetic observables in the SU($2$) flavor group are presented in Appendix~\ref{s:su2}.
In Appendix~\ref{s:coarse} we determine the $\cO(a)$ corrections in an alternative power
counting scheme for coarser lattices where $\e\sim a\L_\chi$. 
A conclusion appears in Section~\ref{sec:conclusions}.

\section{\label{sec:PQCPT}\PQCPT\ at finite lattice spacing}
In partially quenched QCD (PQQCD)~%
\cite{Sharpe:2000bn,Sharpe:2001fh,Sharpe:2000bc,Sharpe:1999kj,Golterman:1998st,Sharpe:1997by,Bernard:1994sv,Shoresh:2001ha} 
the quark part of the Symanzik Lagrangian~%
\cite{Symanzik:1983dc,Symanzik:1983gh} to $\cO(a)$ is written as 
\begin{equation}\label{eqn:LPQQCD}
  {\cal L}
  =
  \ol Q \, (i\Dslash-m_Q) \, Q
  + a \, \ol Q \, \sigma^{\mu \nu} G_{\mu \nu} \,  c_Q \, Q	
,\end{equation}
where the second term, the Pauli-term, breaks chiral symmetry in the same way as the quark mass term.
Here, the nine quarks of PQQCD are in the fundamental representation of
the graded group $SU(6|3)$%
~\cite{BahaBalantekin:1981kt,BahaBalantekin:1981qy,BahaBalantekin:1982bk}
and appear in the vector
\begin{equation}
  Q=(u,d,s,j,l,r,\tilde{u},\tilde{d},\tilde{s})
\end{equation}
that obeys the graded equal-time commutation relation
\begin{equation} \label{eqn:commutation}
  Q^\a_i({\bf x}){Q^\b_j}^\dagger({\bf y})
  -(-1)^{\eta_i \eta_j}{Q^\b_j}^\dagger({\bf y})Q^\a_i({\bf x})
  =
  \d^{\a\b}\d_{ij}\d^3({\bf x}-{\bf y})
,\end{equation}
where $\a$ and $\b$ are spin, and $i$ and $j$ are flavor indices.
The remaining graded equal-time commutation relations can be written analogously.
The different statistics for fermionic and bosonic quarks are incorporated in the grading factor 
\begin{equation}
   \eta_k
   = \left\{ 
       \begin{array}{cl}
         1 & \text{for } k=1,2,3,4,5,6 \\
         0 & \text{for } k=7,8,9
       \end{array}
     \right.
.\end{equation}
The quark mass matrix is given by 
\begin{equation}
  m_Q=\text{diag}(m_u,m_d,m_s,m_j,m_l,m_r,m_u,m_d,m_s),
\end{equation}
while the Sheikholeslami-Wohlert (SW)~\cite{Sheikholeslami:1985ij} 
coefficient matrix for mixed actions reads
\begin{equation} \label{eqn:sw}
c_Q = \text{diag}
(c^{v},c^{v},c^{v},c^{s},c^{s},c^{s},c^{v},c^{v},c^{v}).
\end{equation}
If the quark $Q_i$ is a Wilson fermion~\cite{Wilson:1974sk}, 
then $(c_Q)_i = c_{\text{sw}}$.
Alternately, if $Q_i$ is of the Ginsparg-Wilson variety~\cite{Ginsparg:1982bj}
(e.g., Kaplan fermions~\cite{Kaplan:1992bt} or 
overlap fermions~\cite{Narayanan:1993ss}), 
then $(c_Q)_i = 0$. Since one expects
simulations to be performed with valence quarks that are all of the same species as well as sea quarks 
all of the same species, we have labeled the SW coefficients in Eq.~\eqref{eqn:sw}
by valence (v) and sea (s) instead of flavor. 
In the limit $m_j=m_u$, $m_l=m_d$, and $m_r=m_s$ one recovers QCD at $\cO(a)$.

The light quark electric charge matrix $\cQ$ is not uniquely
defined in PQQCD~\cite{Golterman:2001yv}.  
By imposing the charge matrix $\cQ$ to have vanishing supertrace, 
no new operators
involving the singlet component are introduced.
This can be accomplished 
by~\cite{Chen:2001yi}
\begin{equation} \label{eqn:charge}
  \cQ
  =
  \diag
  \left(
    \frac{2}{3},-\frac{1}{3},-\frac{1}{3},q_j,q_l,q_r,q_j,q_l,q_r
  \right)
.\end{equation}

In addition to the SW term in Eq.~\eqref{eqn:LPQQCD}, the vector-current operator 
of PQQCD also receives corrections at $\cO(a)$. There are three operator 
structures to consider~\cite{Capitani:2000xi}
\begin{align} \label{eqn:vectora}
\cO^\mu_0 &=  a \, \ol Q \, \cQ \, c_{0} m_Q  \, \gamma^\mu  \, Q
\notag \\
\cO^\mu_1 &=  a \, \ol Q \, \cQ \, c_{1} \left(i \tensor D{}^\mu \right) Q
\notag \\
\cO^{\mu}_2 &= a \, D_\nu \left( \ol Q \, \cQ \, c_{2} \, \sigma^{\mu \nu} \, Q \right)
,\end{align}
where $\tensor D{}^\mu = \loarrow D{}^\mu - \roarrow D{}^\mu$ and $D^\mu$ is the gauge 
covariant derivative. The form of the matrices $c_0$, $c_1$, and $c_2$ in PQQCD is 
\begin{equation}
c_j = \text{diag}
\left(c^{v}_{j},c^{v}_{j},c^{v}_j,c^{s}_j,c^{s}_j,c^{s}_j,c^{v}_j,c^{v}_j,c^{v}_j
\right)
,\end{equation}
where $c_j^{v}$ and $c_j^{s}$ are the coefficients of the vector-current
correction operator $\cO_j^\mu$ for valence and sea quarks respectively.
If the vector-current operator is $\cO(a)$ improved in the valence (sea)
sector, then $c_j^v = 0 $ ( $c_j^s = 0 $ ).
The operator $\cO_0^\mu$, which corresponds to a renormalization of the vector current, 
contains a factor of $a \, m_Q$ that renders it $\cO(\e^4)$. Thus contributions to 
electromagnetic observables from $\cO_0^\mu$ are neglected below.
The equations of motion which follow from Eq.~\eqref{eqn:LPQQCD}
can be used to show that the operator $\cO^\mu_2$ is redundant up to $\cO(a^2)$ corrections.
Therefore, we need not consider $\cO^\mu_2$. For ease we define the matrix product $c_{1,\cQ} = \cQ c_1$.

\subsection{Mesons}
For massless quarks at zero lattice spacing,
the Lagrangian in Eq.~(\ref{eqn:LPQQCD}) exhibits a graded symmetry
$SU(6|3)_L \otimes SU(6|3)_R \otimes U(1)_V$ that is assumed 
to be spontaneously broken down to $SU(6|3)_V \otimes U(1)_V$. 
The low-energy effective theory of PQQCD that results from 
expanding about the physical vacuum state is \PQCPT.
The emerging 80~pseudo-Goldstone mesons 
can be described at $\cO(\e^2)$ by a Lagrangian 
which accounts  now for the two sources of explicit chiral symmetry breaking~%
\cite{Sharpe:1998xm,Rupak:2002sm,Bar:2002nr}
\begin{equation}\label{eqn:Lchi}
  {\cal L} =
  \frac{f^2}{8}
    \str\left(D^\mu\Sigma^\dagger D_\mu\Sigma\right)
    + \l_m\,\str\left(m_Q\Sigma+m_Q^\dagger\Sigma^\dagger\right)
    + a \l_a\,\str\left(c_Q\Sigma+c_Q^\dagger\Sigma^\dagger\right)
    + \a\partial^\mu\Phi_0\partial_\mu\Phi_0
    - \mu_0^2\Phi_0^2
\end{equation}
where
\begin{equation} \label{eqn:Sigma}
  \Sigma=\exp\left(\frac{2i\Phi}{f}\right)
  = \xi^2
,\end{equation}
\begin{equation}
  \Phi=
    \left(
      \begin{array}{cc}
        M & \chi^{\dagger} \\ 
        \chi & \tilde{M}
      \end{array}
    \right)
,\end{equation}
$f=132$~MeV,
and the gauge-covariant derivative is
$D_\mu\S=\partial_\mu\S+ie\cA_\mu[\cQ,\S]$.
The str() denotes a graded flavor trace.  
The $M$, $\tilde{M}$, and $\chi$ are matrices
of pseudo-Goldstone bosons 
and pseudo-Goldstone fermions,
see, for example,~\cite{Chen:2001yi}.
Expanding the Lagrangian in \eqref{eqn:Lchi} one finds that
to lowest order mesons with quark content $Q\bar{Q'}$
have mass
\begin{equation}\label{eqn:mqq}
  m_{QQ'}^2=\frac{4}{f^2} \left[ \l_m (m_Q+m_{Q'}) + a \l_a (c_Q + c_{Q'}) \right]
.\end{equation}

The flavor singlet field is $\Phi_0=\str(\Phi)/\sqrt{6}$.
It is rendered heavy by the $U(1)_A$ anomaly
and can be integrated out in \PQCPT,
with its mass $\mu_0$ taken  
on the order of the chiral symmetry breaking scale, 
$\mu_0\to\Lambda_\chi$. In this limit the 
propagator of the flavor singlet field is independent of the
coupling $\a$ and deviates from a simple pole 
form~\cite{Sharpe:2000bn,Sharpe:2001fh}. In \QCPT, the singlet must 
be retained.

\subsection{Baryons}
In PQQCD 
there are baryons 
with quark composition $Q_iQ_jQ_k$ that can 
contain all three types of quarks. The spin-$1/2$ baryons 
are embedded in the $\bf{240}$-dimensional super-multiplet $\cB_{ijk}$, that
contains the familiar octet baryons, while the spin-$3/2$ baryons
are embedded in the $\bf{138}$-dimensional super-multiplet $\cT^\mu_{ijk}$, that 
contains the familiar decuplet baryons~\cite{Labrenz:1996jy,Chen:2001yi}. 

At leading order in the heavy baryon expansion and at $\cO(a)$, the 
free Lagrangian for the $\cB_{ijk}$ and 
$\cT_{ijk}^\mu$ fields is given by~\cite{Labrenz:1996jy,Beane:2003xv}
\begin{eqnarray} \label{eqn:L}
  {\mathcal L}
  &=&
  i\left(\ol\cB v\cdot{\mathcal D}\cB\right)
  +2\a_M\left(\ol\cB \cB{\mathcal M}_+\right)
  +2\b_M\left(\ol\cB {\mathcal M}_+\cB\right)
  +2\sigma_M\left(\ol\cB\cB\right)\str\left({\mathcal M}_+\right)
                              \nonumber \\
  &&+2\a_A\left(\ol\cB \cB{\mathcal A}_+\right)
  +2\b_A\left(\ol\cB {\mathcal A}_+\cB\right)
  +2\sigma_A\left(\ol\cB\cB\right)\str\left({\mathcal A}_+\right)
                              \nonumber \\
  &&-i\left(\ol\cT^\mu v\cdot{\mathcal D}\cT_\mu\right)
  +\D\left(\ol\cT^\mu\cT_\mu\right)
  +2\g_M\left(\ol\cT^\mu {\mathcal M}_+\cT_\mu\right)
  -2\ol\sigma_M\left(\ol\cT^\mu\cT_\mu\right)\str\left({\mathcal M}_+\right) 
				\nonumber \\
  &&+2\g_A\left(\ol\cT^\mu {\mathcal A}_+\cT_\mu\right)
  -2\ol\sigma_A\left(\ol\cT^\mu\cT_\mu\right)\str\left({\mathcal A}_+\right) 
,\end{eqnarray}
where 
${\mathcal M}_+
  =\frac{1}{2}\left(\xi^\dagger m_Q \xi^\dagger+\xi m_Q \xi\right)$
and $\mathcal{A}_+ = \frac{1}{2} a \left(\xi^\dagger c_Q \xi^\dagger+\xi c_Q \xi\right)$. 
Here $\D$ is the mass splitting between the $\bf{240}$ and $\bf{138}$ in the chiral limit.
The parenthesis notation used in Eq.~\eqref{eqn:L} is defined in~\cite{Labrenz:1996jy}
so that the contraction of flavor indices maintains proper transformations under chiral rotations.
Notice that the presence of the chiral symmetry breaking SW operator in Eq.~\eqref{eqn:LPQQCD} 
has lead to new $\cO(a)$ operators (and new constants $\a_A$, $\b_A$, $\sigma_A$, $\g_A$, 
and $\ol\sigma_A$) in Eq.~\eqref{eqn:L}.
The Lagrangian describing the interactions of the $\cB_{ijk}$ 
and $\cT^\mu_{ijk}$ with the pseudo-Goldstone mesons is
\begin{equation} \label{eqn:Linteract}
  {\cal L} =   
	  2 \a \left(\ol \cB S^\mu \cB A_\mu \right)
	+ 2 \b \left(\ol \cB S^\mu A_\mu \cB \right)
	+ 2{\mathcal H}\left(\ol{\cT}^\nu S^\mu A_\mu \cT_\nu\right) 
    	+ \sqrt{\frac{3}{2}}\cC
  		\left[
    			\left(\ol{\cT}^\nu A_\nu \cB\right)+ \left(\ol \cB A_\nu \cT^\nu\right)
  		\right]  
.\end{equation}
The axial-vector and vector meson fields $A^\mu$ and $V^\mu$
are defined by: $ A^\mu=\frac{i}{2}
\left(\xi\partial^\mu\xi^\dagger-\xi^\dagger\partial^\mu\xi\right)$  
and $V^\mu=\frac{1}{2} \left(\xi\partial^\mu\xi^\dagger+\xi^\dagger\partial^\mu\xi\right)$.
The latter appears in  Eq.~\eqref{eqn:L} for the
covariant derivatives of $\cB_{ijk}$ and $\cT_{ijk}$ 
that both have the form
\begin{equation}
  ({\mathcal D}^\mu \cB)_{ijk}
  =
  \partial^\mu \cB_{ijk}
  +(V^\mu)_{il}\cB_{ljk}
  +(-)^{\eta_i(\eta_j+\eta_m)}(V^\mu)_{jm}\cB_{imk}
  +(-)^{(\eta_i+\eta_j)(\eta_k+\eta_n)}(V^\mu)_{kn}\cB_{ijn}
.\end{equation}
The vector $S^\mu$ is the covariant spin operator~\cite{Jenkins:1991jv,Jenkins:1991es,Jenkins:1991ne}.
The parameters $\mathcal{C}$ and $\mathcal{H}$ entering in Eq.~\eqref{eqn:Linteract} are identical to 
those in QCD, while the parameters $\a$ and $\b$ can be related to the familiar constants $D$ and $F$ 
of QCD
\begin{equation}
\a = \frac{2}{3} D + 2 F  \quad\text{and}\quad \b = - \frac{5}{3} D + F
.\end{equation}
In QQCD these identifications cannot be made.

\section{\label{sec:mesons}Octet meson electromagnetic properties}
The electromagnetic form factor $G(q^2)$ of an octet meson 
$\phi$ has the form
\begin{equation}\label{eqn:mesonff}
  \langle\phi(p')|J^\mu|\phi(p)\rangle
  = 
  G(q^2)(p+p')^\mu
\end{equation}
where
$q^\mu=(p'-p)^\mu$. 
At zero momentum transfer $G(0)=Q$, where $Q$ is the charge of $\phi$.
The charge radius $r$ is related to the slope of $G(q^2)$ at $q^2=0$,
namely
\begin{equation}
  <r^2>
  =
  6\frac{d}{dq^2}G(q^2)\Big|_{q^2=0}
.\end{equation}
Recall, at one-loop order in the chiral expansion the charge radii are $\cO(\e^2)$.

There are two finite-$a$ terms in the $\order(\e^4)$ Lagrangian~\cite{Bar:2003mh}
\begin{equation} \label{eqn:L4PQQCD}
  {\cal L} 
  =
  \a_{A,4}\frac{8 a \lambda_a}{f^2}
    \str(D_\mu\Sigma^\dagger D^\mu\Sigma)
    \str(c_Q\Sigma+c_Q^\dagger\Sigma^\dagger)
  +
  \a_{A,5}\frac{8 a \lambda_a}{f^2}  
    \str(D_\mu\Sigma^\dagger D^\mu\Sigma(c_Q\Sigma+c_Q^\dagger\Sigma^\dagger))
\end{equation}
that contribute to meson form factors at tree level.
The new parameters $\a_{A,4}$ and $\a_{A,5}$ in
Eq.~\eqref{eqn:L4PQQCD} are finite lattice spacing analogues of the dimensionless Gasser-Leutwyler
coefficients $\a_4$ and $\a_5$ of \CPT~\cite{Gasser:1985gg}. The above terms contribute 
to meson form factors at $\cO(\e^2)$ but their contributions are independent of $q^2$ and annihilated
by the corresponding wavefunction renormalization,
thus ensuring charge non-renormalization. 

The SW term can potentially contribute at $\cO(\e^2)$ when  
$\cA_+$ is inserted into the kinetic term of the leading-order $\cL$ in 
Eq.~\eqref{eqn:Lchi}. Contributions to form factors from such terms vanish
by charge non-renormalization. 
Insertions of $\cA_+$ into the $\a_9$ term of the Gasser-Leutwyler Lagrangian
produces the $\cO(\e^6)$ terms
\begin{eqnarray} 
\cL &=& 
i m_1 \L_\chi F_{\mu \nu} \, \str 
\left( 
\{\cQ_+,\cA_+\} D^\mu  \Sigma D^\nu \Sigma^\dagger 
+ 
\{\cQ_+,\cA_+\} D^\mu \Sigma^\dagger D^\nu  \Sigma 
\right)
\notag \\
&&+
i m_2 \L_\chi  F_{\mu \nu} \, \str 
\left( 
\cQ_+ D^\mu  \Sigma \cA_+ D^\nu \Sigma^\dagger 
+ 
\cQ_+ D^\mu \Sigma^\dagger \cA_+ D^\nu  \Sigma 
\right)
\notag \\
&&+
i m_3 \L_\chi F_{\mu \nu} \, \str 
\left( 
\cQ_+ D^\mu  \Sigma D^\nu \Sigma^\dagger 
+ 
\cQ_+ D^\mu \Sigma^\dagger D^\nu  \Sigma 
\right) \, \str(\cA_+) 
,\label{eqn:mesonops} \end{eqnarray}
where we have defined $\cQ_+ = \frac{1}{2}\left( \xi^\dagger \cQ \xi^\dagger + \xi \cQ \xi \right)$. 
These terms contribute at $\cO(\e^4)$ to the charge radii 
and can be ignored (see Appendix \ref{s:coarse} for discussion
relating to larger lattice spacings). 

Additionally we must consider the contribution from the vector-current correction 
operator $\cO_1^\mu$ in Eq.~\eqref{eqn:vectora}. In the meson sector, the leading operators $\cO_1^\mu$
in the effective theory can be ascertained by inserting $a \L_\chi c_{1,\cQ}$ in place of $\cQ$ in the 
operators that contribute to form factors. 
The effective field theory operators must also preserve the charge of the meson $\phi$. It is easiest
to embed the operators $\cO_1^\mu$ in a Lagrangian so that electromagnetic gauge invariance is manifest.
To leading order, the contribution from $\cO_1^\mu$ is contained in the term
\begin{equation} \label{eqn:mesonOops}
\cL =
i \a_{A,9} \, a \L_\chi F_{\mu \nu} \, \str 
\left( 
c_{1,\cQ} 
\partial^\mu  \Sigma \partial^\nu \Sigma^\dagger 
+ 
c_{1,\cQ} 
\partial^\mu \Sigma^\dagger \partial^\nu  \Sigma 
\right)  
.\end{equation}
Thus the correction to meson form factors from $\cO^\mu_1$ is at $\cO(\e^4)$. 

The charge radius of the meson $\phi$ to $\cO(\e^2)$ then reads
\begin{equation}
<r^2>
  =  \a_9 \frac{24  Q}{f^2}+
\frac{1}{16 \pi^2 f^2} \sum_X A_X  \log \frac{m_X^2}{\mu^2}  
,\end{equation}
where $X$ corresponds to loop mesons having mass $m_X$ [the masses implicitly include 
the finite lattice spacing corrections given in Eq.~\eqref{eqn:mqq}, otherwise
the expression is identical to the $a=0$ result].
The coefficients $A_X$ in \PQCPT\ appear in Ref.~\cite{Arndt:2003ww}.
In the case of \QCPT, the coefficients $A_X = 0$ for all loop mesons and there are
no additional contributions from the singlet field at this order. Thus  
there is neither quark mass dependence nor lattice spacing dependence in the 
quenched meson charge radii at this order.

\section{\label{sec:octet}Octet baryon electromagnetic properties}
Baryon matrix elements of the electromagnetic current $J^\mu$ can be 
parametrized in terms of the Dirac and Pauli form factors $F_1$ and $F_2$,
respectively, as
\begin{equation}
  \langle\ol B(p') \left|J^\mu\right|B(p)\rangle
  =
  \,\ol u(p')
  \left\{
    v^\mu F_1(q^2)+\frac{[S^\mu,S^\nu]}{M_B}q_\nu F_2(q^2)
  \right\}
  u(p)
\end{equation}
with $q=p'-p$ and $M_B$ is the degenerate octet baryon mass. 
The Dirac form factor is normalized at zero momentum transfer
to the baryon charge: $F_1(0) = Q$.
The electric charge radius $<r_E^2>$, and magnetic moment $\mu$
can be defined in terms of these form factors by
\begin{equation}
<r_E^2>=6 \frac{d F_1(0)}{dq^2} + \frac{3}{2 M_B^2} F_2(0),
\end{equation}
and
\begin{equation}
\mu = F_2(0)
.\end{equation}
Recall, that the one-loop contributions in the chiral expansion to the charge radii are $\cO(\e^2)$, while 
those to the magnetic moments are $\cO(\e)$. 

There are no 
finite-$a$ operators in Eq.~\eqref{eqn:L} that contribute 
to octet baryon form factors.  As in the meson sector, however, the SW term 
could contribute when $\cA_+$ is inserted into the Lagrangian.
Here and henceforth we do not consider these insertions into the kinetic terms in Eq.~\eqref{eqn:L}
because their contributions alter the baryon charges and will be canceled by the appropriate wavefunction 
renormalization. 

The SW term, however, does contribute
when $\cA_+$ is inserted into the charge radius and magnetic moment terms.
For the charge radius, we then have the $\cO(a)$ terms
\begin{eqnarray}
\cL &=&
\frac{b_1}{\L_\chi}  
(-)^{(\eta_i + \eta_j)(\eta_k + \eta_{k^\prime})}
\ol \cB {}^{kji}  
\{\cQ_+,\cA_+\}^{kk^\prime}
\cB^{ijk^\prime} \,
v_{\mu} \partial_\nu F^{\mu \nu}
\notag \\
&&+
\frac{b_2}{\L_\chi}
\ol \cB {}^{kji} \{\cQ_+,\cA_+\}^{ii^\prime} \cB^{i^\prime j k}
\, v_{\mu} \partial_\nu F^{\mu \nu}
\notag \\
&&+
\frac{b_{3}}{\L_\chi} (-)^{\eta_{i'} (\eta_j + \eta_{j'})}
\ol \cB {}^{kji} \cQ_+^{ii'} \cA^{jj'}_+ \cB^{i'j'k} 
v_{\mu} \partial_\nu F^{\mu \nu} 
\notag \\
&&+ 
\frac{b_{4}}{\L_\chi} 
(-)^{\eta_i(\eta_j + \eta_{j'})}
\ol \cB {}^{kji} \cQ_+^{jj'} \cA^{ii'}_+ \cB^{i'j'k}  
\, v_{\mu} \partial_\nu F^{\mu \nu}
\notag \\
&&+
\frac{b_{5}}{\L_\chi}
(-)^{\eta_j \eta_{j'} + 1}
\ol \cB {}^{kji} 
\left(
\cQ_+^{i j'} \cA_+^{j i'} + \cA_+^{i j'} \cQ_+^{j i'}
\right) 
\cB^{i'j'k} 
\, v_{\mu} \partial_\nu  F^{\mu \nu} 
\notag \\
&&+
\frac{1}{\L_\chi}
\left[
b_{6}
\left(
\ol \cB  \cB \cQ_+
\right)
+
b_{7}
\left(
\ol \cB  \cQ_+ \cB
\right)
\right] 
v_{\mu} \partial_\nu
F^{\mu \nu} \, \str(\cA_+) \notag \\
&&+
\frac{b_{8}}{\L_\chi}  
\left(
\ol \cB  \cB \right)
\, v_{\mu} \partial_\nu
F^{\mu \nu} \, \str(\cQ_+ \cA_+)
,\end{eqnarray}
that contribute at $\cO(\e^4)$ to the charge radii and are thus neglected.
Insertions of $\cA_+$ into the magnetic moment terms produce 
\begin{eqnarray} 
\cL &=& 
i b_{1}^\prime
(-)^{(\eta_i + \eta_j)(\eta_k + \eta_{k^\prime})}
\ol \cB {}^{kji}  
[S_\mu,S_\nu]
\{\cQ_+,\cA_+\}^{kk^\prime}
\cB^{ijk^\prime} \,
F^{\mu \nu}
\notag \\
&&+ 
i b_{2}^\prime \,
\ol \cB {}^{kji}  
[S_\mu,S_\nu]
\{\cQ_+,\cA_+\}^{ii^\prime}
\cB^{i^\prime jk} \,
F^{\mu \nu}
\notag \\
&&+
i
b_{3}^\prime (-)^{\eta_{i'} (\eta_j + \eta_{j'})}
\ol \cB {}^{kji} [S_\mu, S_\nu] \cQ_+^{ii'} \cA^{jj'}_+ \cB^{i'j'k} 
F^{\mu \nu} 
\notag \\
&&+ 
i b_{4}^\prime (-)^{\eta_i(\eta_j + \eta_{j'})}
\ol \cB {}^{kji} [S_\mu, S_\nu] \cQ_+^{jj'} \cA^{ii'}_+ \cB^{i'j'k}  
F^{\mu \nu}
\notag \\
&&+
i \, b_{5}^\prime
(-)^{\eta_j \eta_{j'} + 1}
\ol \cB {}^{kji} [S_\mu, S_\nu] 
\left(
\cQ_+^{i j'} \cA_+^{j i'} + \cA_+^{i j'} \cQ_+^{j i'}
\right) 
\cB^{i'j'k} F^{\mu \nu} 
\notag \\
&&+
i 
\left[
b_6^\prime
\left(
\ol \cB [S_\mu,S_\nu] \cB \cQ_+
\right)
+
b_7^\prime
\left(
\ol \cB [S_\mu,S_\nu] \cQ_+ \cB
\right)
\right] F^{\mu \nu} \, \str(\cA_+) \notag \\
&&+
i \, b_8^\prime  
\left(
\ol \cB [S_\mu,S_\nu] \cB \right)
F^{\mu \nu} \, \str(\cQ_+ \cA_+)
\label{eqn:baryonops},\end{eqnarray}
which are $\cO(\e^2)$ corrections to the magnetic moments and can be discarded~\cite{Beane:2003xv}.

Finally we assess the contribution from 
the operator $\cO^\mu_1$ in Eq.~\eqref{eqn:vectora}. As in the meson sector, the charge preserving operators 
can be constructed by the replacement $\cQ \rightarrow a \L_\chi c_{1,\cQ}$ in leading-order terms. Again it is easier to embed 
these operators in $\cL$ so that gauge invariance is transparent. 
For the charge radius, the leading vector-current correction operator is contained in the term
\begin{equation}
\cL =
\frac{a}{\L_\chi} \left[
c_{A,\a} \left(
\ol \cB \cB c_{1,\cQ}
\right)
+
c_{A,\b} \left(
\ol \cB c_{1,\cQ} \cB
\right)
\right] 
v_\mu \partial_\nu  F^{\mu \nu}
,\end{equation} 
which leads to $\cO(\e^4)$ corrections.
For the magnetic moment operator, such a replacement leads to
\begin{equation}\label{eqn:baryonOops}
\cL =
\frac{i a}{2} \left[
\mu_{A,\a} \left(
\ol \cB [S_\mu,S_\nu] \cB c_{1,\cQ}
\right)
+
\mu_{A,\b} \left(
\ol \cB [S_\mu,S_\nu] c_{1,\cQ} \cB
\right)
\right] F^{\mu \nu}
,\end{equation}
and corrections that are of higher order than the one-loop results~\cite{Beane:2003xv}.
See Appendix~\ref{s:coarse} for results in an alternate power counting scheme.

To $\cO(\e^2)$ the baryon charge radii are thus
\begin{eqnarray} 
  <r_E^2>
  &=&
  -\frac{6}{\L_\chi^2}(Qc_-+\a_Dc_+)
  +
  \frac{3}{2M_B^2}(Q\mu_F+\a_D\mu_D)
                    \nonumber \\
  &&
  -
  \frac{1}{ 16\pi^2 f^2}
  \sum_{X}
  \left[ 
    A_X \log\frac{m_X^2}{\mu^2}
    -
    5\,\b_X \log\frac{m_X^2}{\mu^2}
    +
    10\,\b_X'{\mathcal G}(m_X,\D,\mu)
  \right] 
\end{eqnarray}
and the magnetic moments to $\cO(\e)$ read
\begin{equation}
\mu=(Q\,\mu_F+\a_D\,\mu_D)
+ \frac{M_B}{4\pi f^2}\sum_X\left[\b_X m_X
          +\b'_X {\mathcal F}(m_X,\D,\mu)\right]
.\end{equation}
The $a$-dependence is treated as implicit in the meson masses. The \PQCPT\ coefficients
$A_X$, $\b_X$, and $\b_X^{\prime}$ can be found in~\cite{Chen:2001yi,Arndt:2003ww} along
with the functions ${\mathcal F}(m_X,\D,\mu)$ and ${\mathcal G}(m_X,\D,\mu)$. 
The quenched charge radii at $\cO(\e^2)$ are similar in form (although $A_X = 0$ in \QCPT) 
due to the lack of singlet contributions at this order. The \QCPT\ coefficients
$\b^Q_X$ and $\b_X^{Q \prime}$ appear in~\cite{Savage:2001dy,Arndt:2003ww}.
The quenched magnetic moments, however, receive additional contributions from singlet loops.
The relevant formula of~\cite{Savage:2001dy} are not duplicated here in the interests of space
but only need trivial modification by taking into account the $a$-dependence of meson masses.

\section{\label{sec:decuplet}Decuplet baryon electromagnetic properties}
Decuplet matrix elements of the electromagnetic current $J^\rho$ can be parametrized as~\cite{Arndt:2003we}
\begin{equation}
\langle \ol T(p') | J^\rho | T(p) \rangle = -  \ol u_\mu(p') \mathcal{O}^{\mu \rho \nu} u_\nu(p),
\end{equation}
where $u_\mu(p)$ is a Rarita-Schwinger spinor for an on-shell heavy baryon. 
The tensor $\mathcal{O}^{\mu \rho \nu}$ can be parametrized in terms of four independent, 
Lorentz invariant form factors
\begin{equation}
\mathcal{O}^{\mu \rho \nu} = g^{\mu \nu} \left\{ v^\rho F_1(q^2) + \frac{[S^\rho,S^\tau] }{M_B}q_\tau F_2(q^2)  \right\}
+ \frac{q^\mu q^\nu}{(2 M_B)^2} \left\{ v^\rho G_1(q^2) + \frac{[S^\rho,S^\tau]}{M_B} q_\tau G_2(q^2)  \right\},
\end{equation}
where the momentum transfer $q = p' - p$. 
The form factor $F_1(q^2)$ is normalized to the decuplet charge in units of $e$ such that $F_1(0) = Q$.

The conversion from the covariant vertex functions used above to multipole form 
factors for spin-$3/2$ particles is explicated in~\cite{Nozawa:1990gt}.
For our calculations, the charge radius
\begin{equation} \label{eqn:fred}
< r_E^2 >  = 
6 \left\{ \frac{d F_1 (0) }{dq^2} - 
\frac{1}{12 M_B^2} \left[2 Q - 3 F_2(0) - G_1(0) \right]\right\}, 
\end{equation}
the magnetic moment 
\begin{equation}
\mu =  F_{2}(0),
\end{equation}
and the electric quadrupole moment 
\begin{equation}
\mathbb{Q} = -\frac{1}{2} G_1(0).
\end{equation}
The charge radii are $\cO(\e^2)$ at next-to-leading order in the chiral expansion, 
while the magnetic moments are $\cO(\e)$ and the electric quadrupole moments 
are $\cO(\e^0)$.
At one-loop order in the chiral expansion, the magnetic octupole moment is zero.

There are no finite-$a$ operators in Eq.~\eqref{eqn:L} that contribute 
to decuplet baryon form factors.  The SW term can 
potentially contribute when $\cA_+$ is inserted into the Lagrangian.
There are three such terms: 
the charge radius, magnetic moment, and electric quadrupole terms.
Insertions of $\cA_+$ into the charge radius term produces
\begin{eqnarray}
\cL &=&
\frac{d_1}{\L_\chi} 
\ol \cT {}^{\sigma,kji} 
\{ \cQ_+, \cA_+ \}^{ii^\prime} 
\cT_\sigma^{i^\prime jk} \,
v_\mu \partial_\nu F^{\mu \nu}
+ 
\frac{d_2}{\L_\chi} (-)^{\eta_{i'} (\eta_j + \eta_{j'})} \,
\ol \cT {}^{\sigma,kji} \cQ_+^{ii'} \cA_+^{jj'} \cT_\sigma^{i'j'k} \,
v_\mu \partial_\nu F^{\mu \nu}
\notag \\
&&+ 
\frac{ d_3}{\L_\chi}  
\left(
\ol \cT {}^\sigma \cQ_+ \cT_\sigma
\right) 
v_\mu \partial_\nu F^{\mu \nu} \, \str(\cA_+) 
+ 
\frac{d_4}{\L_\chi}  
\left(
\ol \cT {}^\sigma \cT_\sigma 
\right)
v_\mu \partial_\nu F^{\mu \nu} \, \str( \cQ_+ \cA_+)
.\end{eqnarray}
These contribute to decuplet charge radii at $\cO(\e^4)$.
As in the octet sector, insertions of $\cA_+$ into the magnetic moment term, namely
\begin{eqnarray}
\cL &=&
i \, d_1^\prime 
\ol \cT {}^{kji}_\mu 
\{ \cQ_+, \cA_+ \}^{ii^\prime} 
\cT_\nu^{i^\prime jk} 
\, F^{\mu \nu}
+ 
i \, d_2^\prime (-)^{\eta_{i'} (\eta_j + \eta_{j'})} \,
\ol \cT {}^{kji}_\mu \cQ_+^{ii'} \cA_+^{jj'} \cT_\nu^{i'j'k}
F^{\mu \nu}
\notag \\
&&+ 
i \, d_3^\prime  
\left(
\ol \cT_\mu \cQ_+ \cT_\nu
\right) 
F^{\mu \nu} \, \str(\cA_+) 
+ 
i \, d_4^\prime  
\left(
\ol \cT_\mu \cT_\nu 
\right)
F^{\mu \nu} \, \str( \cQ_+ \cA_+)
\label{eqn:decupletops},\end{eqnarray}
produce $\cO(\e^2)$ corrections. 
Likewise, insertions of $\cA_+$ into the electric quadrupole term have the form%
\footnote{
The action of ${}^{\{\ldots\}}$ on Lorentz indices produces the symmetric traceless part
of the tensor, viz., 
$\cO^{\{ \mu \nu \}} = \cO^{\mu \nu} + \cO^{\nu \mu} - \frac{1}{2} g^{\mu \nu} \cO^\a {}_\a$ .
}
\begin{eqnarray}
\cL &=&
\frac{d_1^{\prime\prime}}{\L_\chi} 
\ol \cT {}^{\{\mu,kji} 
\{ \cQ_+, \cA_+ \}^{ii^\prime} 
\cT^{\nu\},i^\prime jk} 
\, v^\a \partial_\mu F_{\nu \a}
+ 
\frac{d_{2}^{\prime\prime}}{\L_\chi} (-)^{\eta_{i'} (\eta_j + \eta_{j'})} \,
\ol \cT {}^{\{ \mu,kji} \cQ_+^{ii'} \cA_+^{jj'} \cT^{\nu\},i'j'k}
v^\a \partial_\mu F_{\nu \a}
\notag \\
&&+ 
\frac{d_{3}^{\prime\prime}}{\L_\chi} 
\left(
\ol \cT {}^{\{\mu} \cQ_+ \cT^{\nu\}}
\right) 
v^\a \partial_\mu F_{\nu \a} \, \str(\cA_+) 
+ 
\frac{d_{4}^{\prime\prime}}{\L_\chi}  
\left(
\ol \cT {}^{\{\mu} \cT^{\nu\}} 
\right)
v^\a \partial_\mu F_{\nu \a} \, \str( \cQ_+ \cA_+)
,\end{eqnarray}
and produce $\cO(\e^2)$ corrections.
All of these corrections are of higher order than the one-loop results.

Finally we assess the contribution from 
the operator $\cO^\mu_1$ in Eq.~\eqref{eqn:vectora}. The effective operators 
can be constructed by replacing $\cQ$ by $a \L_\chi c_{1,\cQ}$ in leading-order terms. 
Embedding these terms in a Lagrangian, we have
\begin{equation} \label{eqn:decupletOops}
\cL = 
\frac{3 a \, c_{A,c} }{\L_\chi} 
\left(
\ol \cT {}^\sigma c_{1,\cQ} \cT_\sigma
\right) 
v_\mu \partial_\nu F^{\mu \nu}
+
3 i a
\,  
\mu_{A,c}
\left(
\ol \cT_\mu c_{1,\cQ} \cT_\nu
\right) F^{\mu \nu}
- 
\frac{3 a \, \mathbb{Q}_{A,c}}{\L_\chi}
\left(
\ol \cT {}^{\{\mu} c_{1,\cQ} \cT^{\nu \}}
\right)
v^\a \partial_\mu F_{\nu \a}
.\end{equation} 
Each of these terms leads to corrections of higher order than the one-loop 
results and can be dropped.
Thus at this order the only finite lattice spacing corrections to decuplet
electromagnetic properties appear in the meson masses. For reference, the expressions are
\begin{eqnarray} \label{eqn:r_E}
<r_E^2> & = & Q \left( \frac{2 \mu_c - 1}{M_B^2} + \frac{\mathbb{Q}_c + 6 c_c}{\L_\chi^2}  \right) 
\nonumber \\
& & 
- \frac{1}{3} \, \frac{9 + 5 \mathcal{C}^2}{16 \pi^2 f^2}  \sum_X  A_X  \log \frac{m_X^2}{\mu^2} 
 - \frac{25}{27}\, \frac{\mathcal{H}^2}{16 \pi^2 f^2} \sum_X A_X  \mathcal{G}(m_X,\D,\mu), \\
\mu & = & 2 \mu_c Q - \frac{ M_B \mathcal{H}^2}{36 \pi f^2} \sum_X A_X \mathcal{F}(m_X, \D,\mu) 
- \frac{\mathcal{C}^2 M_B}{8 \pi f^2} \sum_X A_X m_X,
\end{eqnarray}
and 
\begin{eqnarray}
\mathbb{Q} & = & -2 Q \left( \mu_c + \mathbb{Q}_c \frac{2 M_B^2}{\L_\chi^2}  \right) + \frac{M_B^2 \mathcal{C}^2}{24 \pi^2 f^2} 
\sum_X A_X \log \frac{m_X^2}{\mu^2} 
- \frac{M_B^2 \mathcal{H}^2}{54 \pi^2 f^2} \sum_X A_X \mathcal{G}(m_X,\D,\mu).
\nonumber \\
\end{eqnarray}
The coefficients $A_X$ are tabulated in~\cite{Arndt:2003we}. 
Extending the result to \QCPT, where $A_X = 0$, one must include additional
contributions from singlet loops. With finite lattice spacing corrections, the 
expressions are identical to those in~\cite{Arndt:2003we} except with masses 
given by Eq.~\eqref{eqn:mqq}. Thus for brevity we do not reproduce them here.

\section{\label{sec:trans}Decuplet to octet electromagnetic transitions}
The decuplet to octet matrix elements of the electromagnetic current $J^\mu$ appear as~\cite{Arndt:2003vd}
\begin{equation}
\langle \ol  B(p) | J^\mu | T(p^\prime) \rangle = \ol u(p) \mathcal{O}^{\mu \b} u_\b(p^\prime),
\end{equation}
where the tensor $\mathcal{O}^{\mu \b}$ can be parametrized in terms of three independent, 
Lorentz invariant form factors
\begin{eqnarray}
\mathcal{O}^{\mu \b} &=& \left( q \cdot S \,  g^{\mu \beta} - S^\mu q^\b  \right)  \frac{G_1(q^2)}{M_B} 
+ \left( q \cdot v \, g^{\mu \beta} - v^\mu q^\b  \right) q \cdot S
\frac{G_2(q^2)}{(2 M_B)^2}  \notag \\
&& + \left( q^2 \, g^{\mu \b} - q^\mu q^\b \right) S \cdot q  \frac{G_3(q^2)}{4 M_B^2 \D} 
\end{eqnarray}
where the photon momentum $q = p^\prime - p$. At next-to-leading order in the chiral expansion, we recall that 
$G_1(q^2)$ is $\cO(\e)$ while $G_2(q^2)$ and $G_3(q^2)$ are $\cO(\e^0)$.
The conversion of these vertex covariant functions to multipole
form factors is detailed in~\cite{Jones:1973ky}. 
The multipole moments up to $\cO(\e)$ are%
\footnote{
Here, we count $\e \sim \D/M_B$.
} 
\begin{align}
G_{M1}(0) & = \left( \frac{2}{3} - \frac{\D}{6 M_B} \right) G_1(0) + \frac{\D}{12 M_B} G_2(0) \\
G_{E2}(0) & = \frac{\D}{6 M_B} G_1(0) + \frac{\D}{12 M_B} G_2(0) \\
G_{C2}(0) & = \left( \frac{1}{3} + \frac{\D}{6 M_B} \right) G_1(0) 
	+ \left(\frac{1}{6} + \frac{\D}{6 M_B} \right) G_2(0) + \frac{1}{6} G_3(0)
.\end{align}

There are no new finite-$a$ operators in Eq.~\eqref{eqn:L} that contribute 
to decuplet to octet transition form factors.
Insertion of $\cA_+$ into leading-order transition terms leads to corrections of $\cO(\e^2)$ or smaller.
For completeness the terms are: 
\begin{eqnarray}
\cL &=& 
i t_1 
\ol \cB {}^{kji} S_\mu 
\cQ_+^{il} \cA_+^{li^\prime} 
\cT_\nu^{i^\prime jk} \,
 F^{\mu \nu}
+
i t_2 
\ol \cB {}^{kji} S_\mu 
\cA_+^{il} \cQ_+^{li^\prime} 
\cT_\nu^{i^\prime jk} \,
F^{\mu \nu}
\notag \\
&&+
i t_3  (-)^{ \eta_{i'} ( \eta_j + \eta_{j'} ) } 
\, 
\ol \cB {}^{kji} 
S_\mu \cQ_+^{ii'}  \cA_{+}^{jj'} 
\cT_{\nu}^{i'j'k} 
F^{\mu \nu}
+
i t_4 
(-)^{ \eta_{i'} ( \eta_j + \eta_{j'} ) } 
\ol \cB {}^{kji} 
S_\mu 
\cA_+^{ii'} \cQ_+^{jj'}
\cT_\nu^{i'j'k}
F^{\mu \nu}
\notag \\
&&+
i t_5 
\left(
\ol \cB S_\mu \cQ_+ \cT_\nu
\right) 
F^{\mu \nu} \, \str(\cA_+)
\label{eqn:transops},\end{eqnarray}
for the magnetic dipole transition; and
\begin{eqnarray}
\cL &=& 
\frac{t_1^\prime}{\L_\chi} 
\ol \cB {}^{kji} S^{\{\mu} 
\cQ_+^{il} \cA_+^{li^\prime} 
\cT^{\nu\},i^\prime jk}
\, v^\a \partial_\mu F_{\nu \a}
+
\frac{t_2^\prime}{\L_\chi} 
\ol \cB {}^{kji} S^{\{\mu} 
\cA_+^{il} \cQ_+^{li^\prime} 
\cT^{\nu\},i^\prime jk}
\, v^\a \partial_\mu F_{\nu \a}
\notag \\
&&+
\frac{t_3^\prime}{\L_\chi}
(-)^{ \eta_{i'} ( \eta_j + \eta_{j'} ) } 
\, 
\ol \cB {}^{kji} 
S^{\{\mu} \cQ_+^{ii'}  \cA_{+}^{jj'} 
\cT^{\nu \},i'j'k} 
\, v^\a \partial_\mu F_{\nu \a}
\notag \\
&&+
\frac{t_{4}^\prime}{\L_\chi}
(-)^{ \eta_{i'} ( \eta_j + \eta_{j'} ) } 
\ol \cB {}^{kji} 
S^{\{\mu} 
\cA_+^{ii'} \cQ_+^{jj'}
\cT^{\nu\},i'j'k}
\, v^\a \partial_\mu F_{\nu \a}
\notag \\
&&+
\frac{t_{5}^\prime}{\L_\chi} 
\left(
\ol \cB S^{\{\mu} \cQ_+ \cT^{\nu\}}
\right) 
\, v^\a \partial_\mu F_{\nu \a}
\, \str(\cA_+)
,\end{eqnarray}
for the quadrupole transition.
Finally, insertion of $\cA_+$ into the \PQCPT\ term proportional to
$i(\ol \cB S^\mu \cQ \cT^\nu)\partial^\a\partial_\mu F_{\nu\a}$
leads to
\begin{eqnarray}
\cL &=& 
\frac{i t_{1}^{\prime\prime}}{\L_\chi^2} 
\ol \cB {}^{kji} S_\mu 
\cQ_+^{il} \cA_+^{li^\prime} 
\cT_\nu^{i^\prime jk}
\, \partial^\a \partial^\mu F^\nu{}_{\a}
+
\frac{i t_{2}^{\prime\prime}}{\L_\chi^2} 
\ol \cB^{kji} S_\mu 
\cA_+^{il} \cQ_+^{li^\prime} 
\cT_\nu^{i^\prime jk}
\, \partial^\a \partial^\mu F^\nu{}_{\a}
\notag \\
&&+
\frac{i t_{3}^{\prime\prime}}{\L_\chi^2}  
(-)^{ \eta_{i'} ( \eta_j + \eta_{j'} ) } 
\, 
\ol \cB {}^{kji} 
S_\mu \cQ_+^{ii'}  \cA_{+}^{jj'} 
\cT_{\nu}^{i'j'k} 
\, \partial^\a \partial^\mu F^\nu{}_{\a}
\notag \\
&&+
\frac{i t_{4}^{\prime\prime}}{\L_\chi^2} 
(-)^{ \eta_{i'} ( \eta_j + \eta_{j'} ) } 
\ol \cB {}^{kji} 
S_\mu 
\cA_+^{ii'} \cQ_+^{jj'}
\cT_\nu^{i'j'k}
\,  \partial^\a \partial^\mu F^\nu{}_{\a} 
\notag \\
&&+
\frac{i t_{5}^{\prime\prime}}{\L_\chi^2} 
\left(
\ol \cB S_\mu \cQ_+ \cT_\nu
\right) 
\, \partial^\a \partial^\mu F^\nu{}_{\a}
\, \str(\cA_+)
,\end{eqnarray}
for the Coulomb quadrupole transition.

Similarly, constructing $\cO_1^\mu$ in the effective theory by
replacing $\cQ$ with $a \L_\chi c_{1,\cQ}$ in the transition operators leads to terms of at least $\cO(\e^2)$
which are contained in the terms
\begin{eqnarray}
\cL &=&
i a \, \mu_{A,T} \sqrt{\frac{3}{8}} 
\left(
\ol \cB S_\mu c_{1,\cQ} \cT_\nu
\right) F^{\mu \nu}
+
\frac{a \, \mathbb{Q}_{A,T}}{\L_\chi} \sqrt{\frac{3}{2}} 
\left(
\ol \cB S^{\{\mu} c_{1,\cQ} \cT^{\nu\}}
\right) 
v^\a \partial_\mu F_{\nu \a}
\notag \\
&&+ 
\frac{i a \, c_{A,T}}{\L_\chi^2} \sqrt{\frac{3}{2}} 
\left(
\ol \cB S^\mu c_{1,\cQ} \cT^\nu
\right)
\partial^\a \partial_\mu F_{\nu \a}
\label{eqn:transOops}.\end{eqnarray} 
All of these corrections from effective $\cO_1^\mu$ operators
are of higher order than the one-loop results.
Thus at this order, the only finite lattice spacing corrections to the transition moments appear in the meson masses. 
For reference the expressions are
\begin{eqnarray}
G_1(0) 
&=& 
\frac{\mu_T}{2} \a_T 
-
4\pi \cH \cC \frac{M_B}{\L_\chi^2} 
\sum_X \b_X^T \int_0^1 dx \left( 1 - \frac{x}{3} \right) \mathcal{F}(m_X,x\D,\mu) \notag \\
&& +  
4\pi \cC (D - F) \frac{M_B}{\L_\chi^2} 
\sum_X \b_X^B \int_0^1 dx (1 - x) \mathcal{F}(m_X,-x\D,\mu)
,\end{eqnarray} 
\begin{eqnarray}
G_2(0) 
&=& 
\frac{M_B^2}{\L_\chi^2} \Bigg\{ 
- 4 \mathbb{Q}_T \a_T  
+ 
16 \cH \cC \sum_X \b_X^T \int_0^1 dx \frac{x(1-x)}{3} \mathcal{G}(m_X,x\D,\mu) \notag \\
&& -  
16 \cC (D - F) \sum_X \b_X^B \int_0^1 dx \, x (1 - x) \mathcal{G}(m_X,-x\D,\mu) \Bigg\}
,\end{eqnarray} 
and
\begin{eqnarray}
G_3(0) 
&=& 
-16 \frac{M_B^2}{\L_\chi^2} \sum_X 
\int_0^1 dx \; x(1-x) \left( x - \frac{1}{2}\right) \frac{\D m_X}{m_X^2 - x^2 \D^2} \notag \\ 
&& \times
\left[ \frac{1}{3} \cH \cC \, \b_X^T \, \cR \left( \frac{x \D}{m_X} \right) 
+ 
\cC (D - F) \, \b_X^B \, \cR \left(- \frac{x \D}{m_X} \right)   \right] 
.\end{eqnarray}
The coefficients $\b_B$ and $\b_T$ are tabulated in~\cite{Arndt:2003vd} along with the function $\mathcal{R}(x)$. 
Extending the result to \QCPT, where the coefficients are replaced with their quenched counterparts 
$\b_B^Q$ and $\b_T^Q$, one must include additional
contributions from singlet loops. With finite lattice spacing corrections the 
expressions are identical to those in~\cite{Arndt:2003vd} except with masses 
given by Eq.~\eqref{eqn:mqq}. Thus for brevity we do not reproduce them here.

\section{\label{sec:conclusions}Conclusions}
Above we have calculated the finite lattice spacing corrections to hadronic 
electromagnetic observables in both \QCPT\ and \PQCPT\ for the $SU(3)$ flavor group
in the isospin limit and the $SU(2)$ group with non-degenerate quarks. 
In the power counting scheme of~\cite{Beane:2003xv,Bar:2003mh}, $\cO(a)$ corrections
contribute to electromagnetic observables at higher order
than the one-loop chiral corrections. Thus finite lattice spacing manifests
itself only in the meson masses at this order. 

In practice one should not adhere rigidly to a particular power-counting scheme. 
Each observable should be treated on a case by case basis. The actual size of $a$
and additionally the size of counterterms are needed to address the relevance 
of $\cO(a)$ corrections for real lattice data. 
For this reason we have presented an exhaustive list of $\cO(a)$ operators relevant 
for hadronic electromagnetic properties.
In an alternate power counting for a coarser lattice (as explained in Appendix~\ref{s:coarse}), 
some of the operators listed above
contribute at the same order as the one-loop results in the chiral expansion. 
The corrections detailed in Appendix~\ref{s:coarse} 
in the baryon sector are also necessary if one goes beyond the heavy baryon limit and includes 
recoil terms to the Lagrangian, even in the original power counting of Eq.~\eqref{eqn:pc}.

Knowledge of the low-energy behavior of PQQCD at finite lattice spacing 
is crucial to extrapolate lattice calculations from 
the quark masses used on a finite lattice to the physical world.
The formal behavior of the PQQCD electromagnetic observables 
in the chiral limit has the same form as in QCD.
Moreover, there is 
a well-defined connection to QCD and one can
reliably extrapolate lattice results down to the 
quark masses of reality. For simulations using unimproved lattice 
actions (with Wilson quarks or mixed quarks), our results
will aid in the continuum extrapolation and will help lattice simulations
make contact with real-world data.

\begin{acknowledgments}
We thank Gautam Rupak and  Ruth Van de Water
for helpful discussions,  and 
Martin Savage and Steve Sharpe
for critical comments on the manuscript.
This work is supported in part by the U.S.\ Department of Energy
under Grant No.\ DE-FG03-97ER4014. 
\end{acknowledgments}

\appendix 
\section{\label{s:su2} Hadronic electromagnetic properties in $SU(2)$ flavor
         with non-degenerate quarks at finite lattice spacing}
In this Appendix, we consider the case of $SU(2)$ 
flavor PQQCD%
\footnote{
For brevity we refer to $SU(4|2)$ PQQCD as $SU(2)$. 
The distinction will always be clear.}
 and summarize the changes needed to determine 
finite lattice spacing corrections to the electromagnetic properties 
of hadrons considered above.
For the two flavor case, we keep the up and down valence quark masses 
non-degenerate and similarly for the sea-quarks. 
Thus the quark mass matrix reads
\begin{equation}
m_Q^{SU(2)} = \diag(m_u, m_d, m_j, m_l, m_u, m_d),
\end{equation}
while the SW matrix is 
\begin{equation}
c_Q^{SU(2)} = \diag(c^v, c^v, c^s, c^s, c^v, c^v)
.\end{equation}

Defining ghost and sea quark charges is constrained only by the 
restriction that QCD be recovered
in the limit of appropriately degenerate quark masses. 
Thus the most general form of the charge matrix is
\begin{equation}
  \cQ^{SU(2)} 
  = \diag\left(\frac{2}{3},-\frac{1}{3},q_j,q_l,q_j,q_l \right) 
,\end{equation}
which is not supertraceless.
Analogous to the three flavor case, the vector-current will 
receive $\cO(a)$ corrections from the operators in Eq.~\eqref{eqn:vectora} of 
which only the operator $\cO_1^\mu$ is relevant. The coefficient matrix associated with 
this operator is 
\begin{equation}
c_1^{SU(2)} = \diag(c_1^v, c_1^v, c_1^s, c_1^s, c_1^v, c_1^v)
.\end{equation}

The $\cO(a)$ operators listed above in Sections \ref{sec:mesons}--\ref{sec:trans} are the same for the $SU(2)$ flavor group, however, the coefficients
have different numerical values. Additionally there are operators 
involving $\str(\cQ_+^{SU(2)})$. These are listed for each electromagnetic observable below.

\subsection*{Octet mesons}
In the meson sector, one has the additional term
\begin{eqnarray}
\cL &=& 
i m_4 \L_\chi F_{\mu \nu} \, \str 
\left( 
\cA_+ D^\mu  \Sigma D^\nu \Sigma^\dagger 
+ 
\cA_+ D^\mu \Sigma^\dagger D^\nu  \Sigma 
\right) \, \str(\cQ_+^{SU(2)}) 
\label{eqn:mesonops2}.\end{eqnarray}

\subsection*{Octet baryons}
In the octet baryon sector, there are terms which originate from $\cA_+$ insertions 
\begin{eqnarray}
\cL &=&
\frac{1}{\L_\chi}
\left[
b_{9}
\left(
\ol \cB  \cB \cA_+
\right)
+
b_{10}
\left(
\ol \cB  \cA_+ \cB
\right)
\right] 
v_{\mu} \partial_\nu
F^{\mu \nu} \, \str(\cQ_+^{SU(2)}) 
\notag \\
&&+
\frac{b_{11}}{\L_\chi}  
\left(
\ol \cB  \cB \right)
\, v_{\mu} \partial_\nu
F^{\mu \nu} \, \str(\cQ_+^{SU(2)} ) \, \str(\cA_+)
\notag \\
&&+ i 
\left[
b_9^\prime
\left(
\ol \cB [S_\mu,S_\nu] \cB \cA_+
\right)
+
b_{10}^\prime
\left(
\ol \cB [S_\mu,S_\nu] \cA_+ \cB
\right)
\right] F^{\mu \nu} \, \str(\cQ_+^{SU(2)}) \notag \\
&&+
i \, b_{11}^\prime  
\left(
\ol \cB [S_\mu,S_\nu] \cB \right)
F^{\mu \nu} \, \str(\cQ_+^{SU(2)})\,\str( \cA_+)
\label{eqn:baryonops2},\end{eqnarray}
and additional vector-current correction operators
\begin{equation}
\cL =
\frac{a \, c_{A,\gamma}}{\L_\chi} 
 \left(
\ol \cB \cB \right)
\, v_{\mu} \partial_\nu
F^{\mu \nu} \,
\str (\cQ^{SU(2)} c_1^{SU(2)})
+
\frac{i a \, \mu_{A,\gamma}}{2} 
\left(
\ol \cB [S_\mu,S_\nu] \cB 
\right)
F^{\mu \nu} \str (\cQ^{SU(2)} c_1^{SU(2)})
\label{eqn:baryonOops2}.\end{equation}

\subsection*{Decuplet baryons}
Next in the decuplet sector there are terms that result from $\cA_+$ insertions
\begin{eqnarray}
\cL &=&
\frac{ d_5}{\L_\chi}  
\left(
\ol \cT {}^\sigma \cA_+ \cT_\sigma
\right) 
v_\mu \partial_\nu F^{\mu \nu} \, \str(\cQ_+^{SU(2)}) 
+ 
\frac{d_6}{\L_\chi}  
\left(
\ol \cT {}^\sigma \cT_\sigma 
\right)
v_\mu \partial_\nu F^{\mu \nu} \, \str( \cQ_+^{SU(2)} ) \, \str( \cA_+)
\notag \\
&&+
i \, d_5^\prime  
\left(
\ol \cT_\mu \cA_+ \cT_\nu
\right) 
F^{\mu \nu} \, \str(\cQ_+^{SU(2)}) 
+ 
i \, d_6^\prime  
\left(
\ol \cT_\mu \cT_\nu 
\right)
F^{\mu \nu} \, \str( \cQ_+^{SU(2)}) \, \str( \cA_+)
\notag \\
&&+
\frac{d_{5}^{\prime\prime}}{\L_\chi} 
\left(
\ol \cT {}^{\{\mu} \cA_+ \cT^{\nu\}}
\right) 
v^\a \partial_\mu F_{\nu \a} \, \str(\cQ_+^{SU(2)}) 
+ 
\frac{d_{6}^{\prime\prime}}{\L_\chi}  
\left(
\ol \cT {}^{\{\mu} \cT^{\nu\}} 
\right)
v^\a \partial_\mu F_{\nu \a} \, \str( \cQ_+^{SU(2)})\, \str(\cA_+)
\notag \\
&&
\label{eqn:decupletops2}\end{eqnarray}
and also further vector-current correction operators
\begin{eqnarray}
\cL &=&
\frac{3 a \, c^\prime_{A,\gamma} }{\L_\chi} 
\left(
\ol \cT {}^\sigma  \cT_\sigma
\right) 
v_\mu \partial_\nu F^{\mu \nu}\,
\str (\cQ^{SU(2)} c_1^{SU(2)})
+
3 i a
\,  
\mu^\prime_{A,\gamma}
\left(
\ol \cT_\mu \cT_\nu
\right) F^{\mu \nu}\,
\str (\cQ^{SU(2)} c_1^{SU(2)})
\notag \\
&&- 
\frac{3 a \, \mathbb{Q}_{A,\gamma}}{\L_\chi}
\left(
\ol \cT {}^{\{\mu} \cT^{\nu \}}
\right)
v^\a \partial_\mu F_{\nu \a}\,
\str (\cQ^{SU(2)} c_1^{SU(2)})
\label{eqn:decupletOops2}.\end{eqnarray}

\subsection*{Baryon transitions}
Finally for the transitions, there are only new $\cA_+$ insertions
\begin{eqnarray}
\cL &=&
i t_6 
\left(
\ol \cB S_\mu \cA_+ \cT_\nu
\right) 
F^{\mu \nu} \, \str(\cQ_+^{SU(2)}) 
+ 
\frac{t_{6}^\prime}{\L_\chi} 
\left(
\ol \cB S^{\{\mu} \cA_+ \cT^{\nu\}}
\right) 
\, v^\a \partial_\mu F_{\nu \a}
\, \str(\cQ_+^{SU(2)}) 
\notag \\
&&+
\frac{i t_{6}^{\prime\prime}}{\L_\chi^2} 
\left(
\ol \cB S_\mu \cA_+ \cT_\nu
\right) 
\, \partial^\a \partial^\mu F^\nu{}_{\a}
\, \str(\cQ_+^{SU(2)})
\label{eqn:transops2}.\end{eqnarray}

For each electromagnetic observable considered above,
contributions from all $\cO(a)$ operators in the effective theory are 
of higher order than the one-loop results in the chiral expansion. 
Thus one need only retain the finite lattice spacing corrections to the meson
masses and use the previously found expressions for electromagnetic
properties in $SU(2)$ \PQCPT~\cite{Beane:2002vq,Beane:2003xv,Arndt:2003ww,Arndt:2003we,Arndt:2003vd}.

\section{\label{s:coarse} Coarse-lattice power counting}
In this Appendix, we detail the $\cO(a)$ corrections to electromagnetic 
properties in an alternate power-counting scheme. 
We imagine a sufficiently cursed lattice, where
$a \L_\chi$ can be treated as $\cO(\e)$, so that%
\footnote{
This power counting coupled with the chiral expansion
is most efficient for valence Ginsparg-Wilson quarks
where $\cO(a)$ corrections vanish.
We thank Gautam Rupak for pointing this out.
}
\begin{equation}
\e^2 \sim 
\begin{cases}
 m_q/\L_\chi, \\
 a^2 \L_\chi^2, \\ 
 p^2/\L_\chi^2
\end{cases}
.\end{equation}
In this case, there are known additional 
$\cO(a^2)$ corrections~\cite{Bar:2003mh} to the meson masses that are now at $\cO(\e^2)$
and must be included in expressions for loop diagrams.
The free Lagrangian for $\cB_{ijk}$ and $\cT^\mu_{ijk}$ fields contains additional terms of $\cO(a^2)$ that
correct the baryon masses,
and modify the kinetic terms.
Potential contributions due to the latter, whatever their form, must be canceled by wavefunction renormalization diagrams.
The only contribution of  $\cO(a^2)$ could come from tree-level electromagnetic terms
but these are necessarily higher order.
Thus in this power counting there are no unknown $\cO(a^2)$ corrections
for electromagnetic properties.

The only possible corrections come from the $\cO(a)$ operators assembled above.
A few of these do contribute at tree level
and are spelled out below.

\subsection*{Octet mesons}
The $\cO(a)$ corrections to the meson form factors are now $\cO(\e^3)$ in the power counting. 
While the meson charge radii at NLO in the chiral expansion are at $\cO(\e^2)$, further corrections in the 
chiral expansion are at $\cO(\e^4)$. Thus one can use the $\cO(a)$ operators 
to completely deduce the charge radii to $\cO(\e^3)$
[apart from $\cO(\e^3)$ corrections to the meson masses]. 
These $\cO(a)$ operators are given in Eqs.~\eqref{eqn:mesonops} and \eqref{eqn:mesonOops} and yield a correction
$\delta < r_E^2 >$ to the meson charge radii of the form
\begin{equation} \label{eqn:mesona}
\delta < r_E^2 > = Q \, \frac{24 a \L_\chi}{f^2}
\left[
c^v ( 2 m_1 + m_2) + 3 c^s m_3 + c_1^v \a_{A,9}
\right]
\end{equation}
Notice that there are no corrections associated with an unimproved current operator
in the sea sector since $c_1^s$ is absent.

In the case of $SU(2)$ flavor, there is an additional contribution from the operator in Eq.~\eqref{eqn:mesonops2}.
At tree level, however, this operator vanishes. The only correction to Eq.~\eqref{eqn:mesona} in changing to $SU(2)$ flavor is
to replace $3 c^s$ with $2 c^s$ which reflects the change in the number of sea quarks.

\subsection*{Octet baryons}
For the octet baryon electromagnetic properties, the $\cO(a)$ corrections to the charge radii are now $\cO(\e^3)$
and can be dropped
as they are the same order as neglected $1/M_B$ corrections.
The magnetic moments, however, do receive corrections from local operators. 
Specifically, the $\cO(a)$ operators which contribute to magnetic moments at $\cO(\e)$ 
are insertion of $\cA_+$ into the magnetic moment operator given in Eq.~\eqref{eqn:baryonops} and $\cO_1^\mu$
corrections given in Eq.~\eqref{eqn:baryonOops}. Calculation of these corrections yields a shift in the magnetic moments
\begin{eqnarray}
\delta \mu &=& a M_B \Bigg\{ 
c^v \left[ 
A \left( b^\prime_1 + \frac{1}{2} b^\prime_4 \right)   
- B \left( 2 b^\prime_2 + b^\prime_3 - b^\prime_5  \right)
\right]
+ 3 c^s \left(\frac{1}{2} A \, b^\prime_6  - B \, b^\prime_7  \right)  
\notag \\
&+& C (c^s - c^v) q_{jlr} \, b^\prime_8 
+ \frac{c_1^v}{2} \left[ \frac{1}{2} \mu_{A,\a} A - \mu_{A,\b} B  \right]
\Bigg\}
,\end{eqnarray}
where $q_{jlr} = q_j + q_l + q_r$.
The coefficients $A$ and $B$ are listed for octet baryons in Table \ref{t:AB}, while $C = 1$ for all octet magnetic moments
and $C = 0$ for the $\L\Sigma^0$ transition moment. 
\begin{table}
\caption{\label{t:AB}The coefficients $A$ and $B$ for the octet baryons.}  
\begin{tabular}{l |  c    |   c  }
	\hline
	\hline
	     & $A$ & $B$  \\
	\hline
	$p$        	& $1$ 			& $0$ \\
	$n$     	& $-\frac{1}{3}$ 	& $-\frac{1}{3}$ \\
	$\Sigma^+$     	& $1$ 			& $0$ \\
	$\Sigma^0$     	& $\frac{1}{6}$ 	& $\frac{1}{6}$ \\
	$\Lambda$     	& $-\frac{1}{6}$ 	& $-\frac{1}{6}$ \\
$\Sigma^0 \Lambda$	& $\frac{1}{2 \sqrt{3}}$ 	& $\frac{1}{2 \sqrt{3}}$ \\
	$\Sigma^-$     	& $-\frac{2}{3}$ 	& $\frac{1}{3}$ \\
	$\Xi^0$     	& $-\frac{1}{3}$ 	& $-\frac{1}{3}$ \\
	$\Xi^-$     	& $-\frac{2}{3}$ 	& $\frac{1}{3}$ \\
	\hline
	\hline
\end{tabular}
\end{table} 
Notice that there are no corrections associated with an unimproved current operator
in the sea sector.

In the case of $SU(2)$ flavor, there are additional contributions given in Eqs.~\eqref{eqn:baryonops2} and \eqref{eqn:baryonOops2}.
For the proton and neutron, we have
\begin{eqnarray}
\delta \mu^{SU(2)} &=& a M_B \Bigg\{ 
c^v \left[
A 
\left( b^\prime_1 + \frac{1}{2} b^\prime_4 \right)   
- 
B 
\left( 2 b^\prime_2 + b^\prime_3 - b^\prime_5  \right)
+ 
\frac{1}{3} \left( b^\prime_9 + b^\prime_{10} \right)
\right]
\notag \\
&+&
2 c^s
\left( \frac{1}{2} A \, b^\prime_6 -  B \, b^\prime_7 + \frac{1}{3} b^\prime_{11} \right)
+ 
\left[  
c^s q_{jl} + c^v \left( \frac{1}{3} - q_{jl} \right)
b^\prime_8
\right]
\notag \\
&+& 
\frac{c_1^v}{2} 
\left[
\frac{1}{2} A \, \mu_{A,\a} - B \, \mu_{A,\b} + \left( \frac{1}{3} - q_{jl} \right) \mu_{A,\gamma}
\right]
+ 
\frac{c_1^s}{2} \, q_{jl} \, \mu_{A,\gamma}
\Bigg\}
,\end{eqnarray}
where $q_{jl} = q_j + q_l$.

\subsection*{Decuplet baryons}
For the decuplet baryon electromagnetic properties in coarse-lattice power counting, the $\cO(a)$ corrections to the charge radii are $\cO(\e^3)$
and the corrections to the electric quadrupole moments are $\cO(\e)$, both of which are higher order than the one-loop results. 
The magnetic moments, however, do 
receive corrections from local operators. Specifically, the $\cO(a)$ operators which contribute to magnetic moments at $\cO(\e)$ 
are $\cA_+$ insertions into the magnetic moment operator given in Eq.~\eqref{eqn:decupletops} and $\cO_1^\mu$
correction operators given in Eq.~\eqref{eqn:decupletOops}. Calculation of these corrections yields a shift in the magnetic moments
\begin{equation}
\delta \mu = 2 a M_B
\left[ 
\frac{1}{3} c^v Q \, ( 2 d_1^\prime + d_2^\prime )  + c^s Q \, d_3^\prime
+ ( c^s - c^v) q_{jlr} d^\prime_4 + c_1^v Q \, \mu_{A,c}
\right]
.\end{equation}
Notice that in $SU(3)$ $\str \cQ=0$, 
hence there is no dependence on $c_1^s$ in the above result.

In the case of $SU(2)$ flavor, there are additional contributions given in Eqs.~\eqref{eqn:decupletops2} and \eqref{eqn:decupletOops2}.
The corrections to the $\D$ quartet magnetic moments are then
\begin{eqnarray}
\delta \mu^{SU(2)} &=& 2 a M_B
\Bigg\{
\frac{1}{3} c^v \left( 2 Q \, d_1^\prime + Q \, d_2^\prime + d_5^\prime  \right)
+ \frac{2}{3} c^s \left( Q \, d_3^\prime + d_6^\prime  \right)
+ \left[ c^s q_{jl} + c^v \left( \frac{1}{3} - q_{jl} \right) \right] d_4^\prime
\notag \\
&+&
c_1^v \left[ Q \, \mu_{A,c} + ( 1 - 3 q_{jl}) \mu^\prime_{A,\gamma} \right]
+
3 c_1^s \, q_{jl} \, \mu^\prime_{A,\gamma}
\Bigg\}
\end{eqnarray}

\subsection*{Baryon transitions}
For the decuplet to octet electromagnetic transitions in coarse-lattice power counting, the $\cO(a)$ corrections to $G_2(0)$ and $G_3(0)$ are $\cO(\e)$
which are of higher order than the one-loop results.  The $G_1(q^2)$ form factor does, however, 
receive corrections from local operators. Specifically these $\cO(a)$ operators which contribute to $G_1(0)$  at $\cO(\e)$ 
are the insertions of $\cA_+$ into the magnetic dipole transition operator given in Eq.~\eqref{eqn:transops} and the vector-current corrections
given in Eq.~\eqref{eqn:transOops}. Calculation of these corrections yields a shift of $G_1(0)$
\begin{equation} \label{eqn:transa}
\delta G_1(0) = a M_B \, \a_T \sqrt{\frac{2}{3}} 
\left\{  
c^v \left( t_1 + t_2 + t_3 - \frac{1}{2} t_4 \right)
+ 
3 c^s \, t_5
+ 
c_1^v \, \mu_{A,T} \sqrt{\frac{3}{8}} 
\right\}
,\end{equation}
where the transition coefficients $\a_T$ appear in \cite{Arndt:2003vd}.
Again, at this order the result is independent of $\cO(a)$ improvement
to the electromagnetic current in the sea sector.
In the case of $SU(2)$ flavor, there is an additional dipole operator given in Eq.~\eqref{eqn:transops2}. At tree level, however, this operator vanishes.
The only correction to Eq.~\eqref{eqn:transa} in changing to $SU(2)$ flavor is
to replace $3 c^s$ with $2 c^s$ which reflects the change in the number of sea quarks.


\begin{thebibliography}{51}
\expandafter\ifx\csname natexlab\endcsname\relax\def\natexlab#1{#1}\fi
\expandafter\ifx\csname bibnamefont\endcsname\relax
  \def\bibnamefont#1{#1}\fi
\expandafter\ifx\csname bibfnamefont\endcsname\relax
  \def\bibfnamefont#1{#1}\fi
\expandafter\ifx\csname citenamefont\endcsname\relax
  \def\citenamefont#1{#1}\fi
\expandafter\ifx\csname url\endcsname\relax
  \def\url#1{\texttt{#1}}\fi
\expandafter\ifx\csname urlprefix\endcsname\relax\def\urlprefix{URL }\fi
\providecommand{\bibinfo}[2]{#2}
\providecommand{\eprint}[2][]{\url{#2}}

\bibitem[{\citenamefont{Morel}(1987)}]{Morel:1987xk}
\bibinfo{author}{\bibfnamefont{A.}~\bibnamefont{Morel}}, \bibinfo{journal}{J.
  Phys. (France)} \textbf{\bibinfo{volume}{48}}, \bibinfo{pages}{1111}
  (\bibinfo{year}{1987}).

\bibitem[{\citenamefont{Sharpe}(1992)}]{Sharpe:1992ft}
\bibinfo{author}{\bibfnamefont{S.~R.} \bibnamefont{Sharpe}},
  \bibinfo{journal}{Phys. Rev.} \textbf{\bibinfo{volume}{D46}},
  \bibinfo{pages}{3146} (\bibinfo{year}{1992}),
  \eprint[http://arXiv.org/abs]{hep-lat/9205020}.

\bibitem[{\citenamefont{Bernard and
  Golterman}(1992{\natexlab{a}})}]{Bernard:1992mk}
\bibinfo{author}{\bibfnamefont{C.~W.} \bibnamefont{Bernard}} \bibnamefont{and}
  \bibinfo{author}{\bibfnamefont{M.~F.~L.} \bibnamefont{Golterman}},
  \bibinfo{journal}{Phys. Rev.} \textbf{\bibinfo{volume}{D46}},
  \bibinfo{pages}{853} (\bibinfo{year}{1992}{\natexlab{a}}),
  \eprint[http://arXiv.org/abs]{hep-lat/9204007}.

\bibitem[{\citenamefont{Bernard and
  Golterman}(1992{\natexlab{b}})}]{Bernard:1992ep}
\bibinfo{author}{\bibfnamefont{C.~W.} \bibnamefont{Bernard}} \bibnamefont{and}
  \bibinfo{author}{\bibfnamefont{M.}~\bibnamefont{Golterman}},
  \bibinfo{journal}{Nucl. Phys. Proc. Suppl.} \textbf{\bibinfo{volume}{26}},
  \bibinfo{pages}{360} (\bibinfo{year}{1992}{\natexlab{b}}).

\bibitem[{\citenamefont{Golterman}(1994)}]{Golterman:1994mk}
\bibinfo{author}{\bibfnamefont{M.~F.~L.} \bibnamefont{Golterman}},
  \bibinfo{journal}{Acta Phys. Polon.} \textbf{\bibinfo{volume}{B25}},
  \bibinfo{pages}{1731} (\bibinfo{year}{1994}),
  \eprint[http://arXiv.org/abs]{hep-lat/9411005}.

\bibitem[{\citenamefont{Labrenz and Sharpe}(1996)}]{Labrenz:1996jy}
\bibinfo{author}{\bibfnamefont{J.~N.} \bibnamefont{Labrenz}} \bibnamefont{and}
  \bibinfo{author}{\bibfnamefont{S.~R.} \bibnamefont{Sharpe}},
  \bibinfo{journal}{Phys. Rev.} \textbf{\bibinfo{volume}{D54}},
  \bibinfo{pages}{4595} (\bibinfo{year}{1996}),
  \eprint[http://arXiv.org/abs]{hep-lat/9605034}.

\bibitem[{\citenamefont{Sharpe and Zhang}(1996)}]{Sharpe:1996qp}
\bibinfo{author}{\bibfnamefont{S.~R.} \bibnamefont{Sharpe}} \bibnamefont{and}
  \bibinfo{author}{\bibfnamefont{Y.}~\bibnamefont{Zhang}},
  \bibinfo{journal}{Phys. Rev.} \textbf{\bibinfo{volume}{D53}},
  \bibinfo{pages}{5125} (\bibinfo{year}{1996}),
  \eprint[http://arXiv.org/abs]{hep-lat/9510037}.

\bibitem[{\citenamefont{Bernard and Golterman}(1994)}]{Bernard:1994sv}
\bibinfo{author}{\bibfnamefont{C.~W.} \bibnamefont{Bernard}} \bibnamefont{and}
  \bibinfo{author}{\bibfnamefont{M.~F.~L.} \bibnamefont{Golterman}},
  \bibinfo{journal}{Phys. Rev.} \textbf{\bibinfo{volume}{D49}},
  \bibinfo{pages}{486} (\bibinfo{year}{1994}),
  \eprint[http://arXiv.org/abs]{hep-lat/9306005}.

\bibitem[{\citenamefont{Sharpe}(1997)}]{Sharpe:1997by}
\bibinfo{author}{\bibfnamefont{S.~R.} \bibnamefont{Sharpe}},
  \bibinfo{journal}{Phys. Rev.} \textbf{\bibinfo{volume}{D56}},
  \bibinfo{pages}{7052} (\bibinfo{year}{1997}),
  \eprint[http://arXiv.org/abs]{hep-lat/9707018}.

\bibitem[{\citenamefont{Golterman and Leung}(1998)}]{Golterman:1998st}
\bibinfo{author}{\bibfnamefont{M.~F.~L.} \bibnamefont{Golterman}}
  \bibnamefont{and} \bibinfo{author}{\bibfnamefont{K.-C.} \bibnamefont{Leung}},
  \bibinfo{journal}{Phys. Rev.} \textbf{\bibinfo{volume}{D57}},
  \bibinfo{pages}{5703} (\bibinfo{year}{1998}),
  \eprint[http://arXiv.org/abs]{hep-lat/9711033}.

\bibitem[{\citenamefont{Sharpe and
  Shoresh}(2000{\natexlab{a}})}]{Sharpe:1999kj}
\bibinfo{author}{\bibfnamefont{S.~R.} \bibnamefont{Sharpe}} \bibnamefont{and}
  \bibinfo{author}{\bibfnamefont{N.}~\bibnamefont{Shoresh}},
  \bibinfo{journal}{Nucl. Phys. Proc. Suppl.} \textbf{\bibinfo{volume}{83}},
  \bibinfo{pages}{968} (\bibinfo{year}{2000}{\natexlab{a}}),
  \eprint[http://arXiv.org/abs]{hep-lat/9909090}.

\bibitem[{\citenamefont{Sharpe and
  Shoresh}(2001{\natexlab{a}})}]{Sharpe:2000bn}
\bibinfo{author}{\bibfnamefont{S.~R.} \bibnamefont{Sharpe}} \bibnamefont{and}
  \bibinfo{author}{\bibfnamefont{N.}~\bibnamefont{Shoresh}},
  \bibinfo{journal}{Int. J. Mod. Phys.} \textbf{\bibinfo{volume}{A16S1C}},
  \bibinfo{pages}{1219} (\bibinfo{year}{2001}{\natexlab{a}}),
  \eprint[http://arXiv.org/abs]{hep-lat/0011089}.

\bibitem[{\citenamefont{Sharpe and
  Shoresh}(2000{\natexlab{b}})}]{Sharpe:2000bc}
\bibinfo{author}{\bibfnamefont{S.~R.} \bibnamefont{Sharpe}} \bibnamefont{and}
  \bibinfo{author}{\bibfnamefont{N.}~\bibnamefont{Shoresh}},
  \bibinfo{journal}{Phys. Rev.} \textbf{\bibinfo{volume}{D62}},
  \bibinfo{pages}{094503} (\bibinfo{year}{2000}{\natexlab{b}}),
  \eprint[http://arXiv.org/abs]{hep-lat/0006017}.

\bibitem[{\citenamefont{Sharpe and
  Shoresh}(2001{\natexlab{b}})}]{Sharpe:2001fh}
\bibinfo{author}{\bibfnamefont{S.~R.} \bibnamefont{Sharpe}} \bibnamefont{and}
  \bibinfo{author}{\bibfnamefont{N.}~\bibnamefont{Shoresh}},
  \bibinfo{journal}{Phys. Rev.} \textbf{\bibinfo{volume}{D64}},
  \bibinfo{pages}{114510} (\bibinfo{year}{2001}{\natexlab{b}}),
  \eprint[http://arXiv.org/abs]{hep-lat/0108003}.

\bibitem[{\citenamefont{Shoresh}(2001)}]{Shoresh:2001ha}
\bibinfo{author}{\bibfnamefont{N.}~\bibnamefont{Shoresh}}
  (\bibinfo{year}{2001}), \bibinfo{note}{{Ph.D.} thesis, University of
  Washington, {UMI}-30-36529}.

\bibitem[{\citenamefont{Sharpe and Van~de Water}(2003)}]{Sharpe:2003vy}
\bibinfo{author}{\bibfnamefont{S.~R.} \bibnamefont{Sharpe}} \bibnamefont{and}
  \bibinfo{author}{\bibfnamefont{R.~S.} \bibnamefont{Van~de Water}}
  (\bibinfo{year}{2003}), \eprint{hep-lat/0310012}.

\bibitem[{\citenamefont{Arndt and Tiburzi}(2003{\natexlab{a}})}]{Arndt:2003ww}
\bibinfo{author}{\bibfnamefont{D.}~\bibnamefont{Arndt}} \bibnamefont{and}
  \bibinfo{author}{\bibfnamefont{B.~C.} \bibnamefont{Tiburzi}},
  \bibinfo{journal}{Phys. Rev.} \textbf{\bibinfo{volume}{D68}},
  \bibinfo{pages}{094501} (\bibinfo{year}{2003}{\natexlab{a}}),
  \eprint{hep-lat/0307003}.

\bibitem[{\citenamefont{Arndt and Tiburzi}(2003{\natexlab{b}})}]{Arndt:2003we}
\bibinfo{author}{\bibfnamefont{D.}~\bibnamefont{Arndt}} \bibnamefont{and}
  \bibinfo{author}{\bibfnamefont{B.~C.} \bibnamefont{Tiburzi}},
  \bibinfo{journal}{Phys. Rev.} \textbf{\bibinfo{volume}{D68}},
  \bibinfo{pages}{114503} (\bibinfo{year}{2003}{\natexlab{b}}),
  \eprint{hep-lat/0308001}.

\bibitem[{\citenamefont{Arndt and Tiburzi}(2004)}]{Arndt:2003vd}
\bibinfo{author}{\bibfnamefont{D.}~\bibnamefont{Arndt}} \bibnamefont{and}
  \bibinfo{author}{\bibfnamefont{B.~C.} \bibnamefont{Tiburzi}},
  \bibinfo{journal}{Phys. Rev.} \textbf{\bibinfo{volume}{D69}},
  \bibinfo{pages}{014501} (\bibinfo{year}{2004}), \eprint{hep-lat/0309013}.

\bibitem[{\citenamefont{Booth}(1995)}]{Booth:1995hx}
\bibinfo{author}{\bibfnamefont{M.~J.} \bibnamefont{Booth}},
  \bibinfo{journal}{Phys. Rev.} \textbf{\bibinfo{volume}{D51}},
  \bibinfo{pages}{2338} (\bibinfo{year}{1995}),
  \eprint[http://arXiv.org/abs]{hep-ph/9411433}.

\bibitem[{\citenamefont{Kim and Kim}(1998)}]{Kim:1998bz}
\bibinfo{author}{\bibfnamefont{M.}~\bibnamefont{Kim}} \bibnamefont{and}
  \bibinfo{author}{\bibfnamefont{S.}~\bibnamefont{Kim}},
  \bibinfo{journal}{Phys. Rev.} \textbf{\bibinfo{volume}{D58}},
  \bibinfo{pages}{074509} (\bibinfo{year}{1998}), \eprint{hep-lat/9608091}.

\bibitem[{\citenamefont{Savage}(2002)}]{Savage:2001dy}
\bibinfo{author}{\bibfnamefont{M.~J.} \bibnamefont{Savage}},
  \bibinfo{journal}{Nucl. Phys.} \textbf{\bibinfo{volume}{A700}},
  \bibinfo{pages}{359} (\bibinfo{year}{2002}), \eprint{nucl-th/0107038}.

\bibitem[{\citenamefont{Arndt}(2003)}]{Arndt:2002ed}
\bibinfo{author}{\bibfnamefont{D.}~\bibnamefont{Arndt}},
  \bibinfo{journal}{Phys. Rev.} \textbf{\bibinfo{volume}{D67}},
  \bibinfo{pages}{074501} (\bibinfo{year}{2003}), \eprint{hep-lat/0210019}.

\bibitem[{\citenamefont{Dong et~al.}(2003)}]{Dong:2003im}
\bibinfo{author}{\bibfnamefont{S.~J.} \bibnamefont{Dong}} \bibnamefont{et~al.}
  (\bibinfo{year}{2003}), \eprint{hep-lat/0304005}.

\bibitem[{\citenamefont{Sharpe and Singleton}(1998)}]{Sharpe:1998xm}
\bibinfo{author}{\bibfnamefont{S.~R.} \bibnamefont{Sharpe}} \bibnamefont{and}
  \bibinfo{author}{\bibfnamefont{J.}~\bibnamefont{Singleton},
  \bibfnamefont{Robert}}, \bibinfo{journal}{Phys. Rev.}
  \textbf{\bibinfo{volume}{D58}}, \bibinfo{pages}{074501}
  (\bibinfo{year}{1998}), \eprint{hep-lat/9804028}.

\bibitem[{\citenamefont{Lee and Sharpe}(1999)}]{Lee:1999zx}
\bibinfo{author}{\bibfnamefont{W.-J.} \bibnamefont{Lee}} \bibnamefont{and}
  \bibinfo{author}{\bibfnamefont{S.~R.} \bibnamefont{Sharpe}},
  \bibinfo{journal}{Phys. Rev.} \textbf{\bibinfo{volume}{D60}},
  \bibinfo{pages}{114503} (\bibinfo{year}{1999}), \eprint{hep-lat/9905023}.

\bibitem[{\citenamefont{Rupak and Shoresh}(2002)}]{Rupak:2002sm}
\bibinfo{author}{\bibfnamefont{G.}~\bibnamefont{Rupak}} \bibnamefont{and}
  \bibinfo{author}{\bibfnamefont{N.}~\bibnamefont{Shoresh}},
  \bibinfo{journal}{Phys. Rev.} \textbf{\bibinfo{volume}{D66}},
  \bibinfo{pages}{054503} (\bibinfo{year}{2002}), \eprint{hep-lat/0201019}.

\bibitem[{\citenamefont{Bar et~al.}(2003{\natexlab{a}})\citenamefont{Bar,
  Rupak, and Shoresh}}]{Bar:2002nr}
\bibinfo{author}{\bibfnamefont{O.}~\bibnamefont{Bar}},
  \bibinfo{author}{\bibfnamefont{G.}~\bibnamefont{Rupak}}, \bibnamefont{and}
  \bibinfo{author}{\bibfnamefont{N.}~\bibnamefont{Shoresh}},
  \bibinfo{journal}{Phys. Rev.} \textbf{\bibinfo{volume}{D67}},
  \bibinfo{pages}{114505} (\bibinfo{year}{2003}{\natexlab{a}}),
  \eprint{hep-lat/0210050}.

\bibitem[{\citenamefont{Bar et~al.}(2003{\natexlab{b}})\citenamefont{Bar,
  Rupak, and Shoresh}}]{Bar:2003mh}
\bibinfo{author}{\bibfnamefont{O.}~\bibnamefont{Bar}},
  \bibinfo{author}{\bibfnamefont{G.}~\bibnamefont{Rupak}}, \bibnamefont{and}
  \bibinfo{author}{\bibfnamefont{N.}~\bibnamefont{Shoresh}}
  (\bibinfo{year}{2003}{\natexlab{b}}), \eprint{hep-lat/0306021}.

\bibitem[{\citenamefont{Aoki}(2003)}]{Aoki:2003yv}
\bibinfo{author}{\bibfnamefont{S.}~\bibnamefont{Aoki}}, \bibinfo{journal}{Phys.
  Rev.} \textbf{\bibinfo{volume}{D68}}, \bibinfo{pages}{054508}
  (\bibinfo{year}{2003}), \eprint{hep-lat/0306027}.

\bibitem[{\citenamefont{Beane and Savage}(2003)}]{Beane:2003xv}
\bibinfo{author}{\bibfnamefont{S.~R.} \bibnamefont{Beane}} \bibnamefont{and}
  \bibinfo{author}{\bibfnamefont{M.~J.} \bibnamefont{Savage}},
  \bibinfo{journal}{Phys. Rev.} \textbf{\bibinfo{volume}{D68}},
  \bibinfo{pages}{114502} (\bibinfo{year}{2003}), \eprint{hep-lat/0306036}.

\bibitem[{\citenamefont{Symanzik}(1983{\natexlab{a}})}]{Symanzik:1983dc}
\bibinfo{author}{\bibfnamefont{K.}~\bibnamefont{Symanzik}},
  \bibinfo{journal}{Nucl. Phys.} \textbf{\bibinfo{volume}{B226}},
  \bibinfo{pages}{187} (\bibinfo{year}{1983}{\natexlab{a}}).

\bibitem[{\citenamefont{Symanzik}(1983{\natexlab{b}})}]{Symanzik:1983gh}
\bibinfo{author}{\bibfnamefont{K.}~\bibnamefont{Symanzik}},
  \bibinfo{journal}{Nucl. Phys.} \textbf{\bibinfo{volume}{B226}},
  \bibinfo{pages}{205} (\bibinfo{year}{1983}{\natexlab{b}}).

\bibitem[{\citenamefont{Balantekin et~al.}(1981)\citenamefont{Balantekin, Bars,
  and Iachello}}]{BahaBalantekin:1981kt}
\bibinfo{author}{\bibfnamefont{A.~B.} \bibnamefont{Balantekin}},
  \bibinfo{author}{\bibfnamefont{I.}~\bibnamefont{Bars}}, \bibnamefont{and}
  \bibinfo{author}{\bibfnamefont{F.}~\bibnamefont{Iachello}},
  \bibinfo{journal}{Phys. Rev. Lett.} \textbf{\bibinfo{volume}{47}},
  \bibinfo{pages}{19} (\bibinfo{year}{1981}).

\bibitem[{\citenamefont{Balantekin and Bars}(1981)}]{BahaBalantekin:1981qy}
\bibinfo{author}{\bibfnamefont{A.~B.} \bibnamefont{Balantekin}}
  \bibnamefont{and} \bibinfo{author}{\bibfnamefont{I.}~\bibnamefont{Bars}},
  \bibinfo{journal}{J. Math. Phys.} \textbf{\bibinfo{volume}{22}},
  \bibinfo{pages}{1149} (\bibinfo{year}{1981}).

\bibitem[{\citenamefont{Balantekin and Bars}(1982)}]{BahaBalantekin:1982bk}
\bibinfo{author}{\bibfnamefont{A.~B.} \bibnamefont{Balantekin}}
  \bibnamefont{and} \bibinfo{author}{\bibfnamefont{I.}~\bibnamefont{Bars}},
  \bibinfo{journal}{J. Math. Phys.} \textbf{\bibinfo{volume}{23}},
  \bibinfo{pages}{1239} (\bibinfo{year}{1982}).

\bibitem[{\citenamefont{Sheikholeslami and
  Wohlert}(1985)}]{Sheikholeslami:1985ij}
\bibinfo{author}{\bibfnamefont{B.}~\bibnamefont{Sheikholeslami}}
  \bibnamefont{and} \bibinfo{author}{\bibfnamefont{R.}~\bibnamefont{Wohlert}},
  \bibinfo{journal}{Nucl. Phys.} \textbf{\bibinfo{volume}{B259}},
  \bibinfo{pages}{572} (\bibinfo{year}{1985}).

\bibitem[{\citenamefont{Wilson}(1974)}]{Wilson:1974sk}
\bibinfo{author}{\bibfnamefont{K.~G.} \bibnamefont{Wilson}},
  \bibinfo{journal}{Phys. Rev.} \textbf{\bibinfo{volume}{D10}},
  \bibinfo{pages}{2445} (\bibinfo{year}{1974}).

\bibitem[{\citenamefont{Ginsparg and Wilson}(1982)}]{Ginsparg:1982bj}
\bibinfo{author}{\bibfnamefont{P.~H.} \bibnamefont{Ginsparg}} \bibnamefont{and}
  \bibinfo{author}{\bibfnamefont{K.~G.} \bibnamefont{Wilson}},
  \bibinfo{journal}{Phys. Rev.} \textbf{\bibinfo{volume}{D25}},
  \bibinfo{pages}{2649} (\bibinfo{year}{1982}).

\bibitem[{\citenamefont{Kaplan}(1992)}]{Kaplan:1992bt}
\bibinfo{author}{\bibfnamefont{D.~B.} \bibnamefont{Kaplan}},
  \bibinfo{journal}{Phys. Lett.} \textbf{\bibinfo{volume}{B288}},
  \bibinfo{pages}{342} (\bibinfo{year}{1992}), \eprint{hep-lat/9206013}.

\bibitem[{\citenamefont{Narayanan and Neuberger}(1993)}]{Narayanan:1993ss}
\bibinfo{author}{\bibfnamefont{R.}~\bibnamefont{Narayanan}} \bibnamefont{and}
  \bibinfo{author}{\bibfnamefont{H.}~\bibnamefont{Neuberger}},
  \bibinfo{journal}{Phys. Rev. Lett.} \textbf{\bibinfo{volume}{71}},
  \bibinfo{pages}{3251} (\bibinfo{year}{1993}), \eprint{hep-lat/9308011}.

\bibitem[{\citenamefont{Golterman and Pallante}(2002)}]{Golterman:2001yv}
\bibinfo{author}{\bibfnamefont{M.}~\bibnamefont{Golterman}} \bibnamefont{and}
  \bibinfo{author}{\bibfnamefont{E.}~\bibnamefont{Pallante}},
  \bibinfo{journal}{Nucl. Phys. Proc. Suppl.} \textbf{\bibinfo{volume}{106}},
  \bibinfo{pages}{335} (\bibinfo{year}{2002}),
  \eprint[http://arXiv.org/abs]{hep-lat/0110183}.

\bibitem[{\citenamefont{Chen and Savage}(2002)}]{Chen:2001yi}
\bibinfo{author}{\bibfnamefont{J.-W.} \bibnamefont{Chen}} \bibnamefont{and}
  \bibinfo{author}{\bibfnamefont{M.~J.} \bibnamefont{Savage}},
  \bibinfo{journal}{Phys. Rev.} \textbf{\bibinfo{volume}{D65}},
  \bibinfo{pages}{094001} (\bibinfo{year}{2002}),
  \eprint[http://arXiv.org/abs]{hep-lat/0111050}.

\bibitem[{\citenamefont{Capitani et~al.}(2001)}]{Capitani:2000xi}
\bibinfo{author}{\bibfnamefont{S.}~\bibnamefont{Capitani}}
  \bibnamefont{et~al.}, \bibinfo{journal}{Nucl. Phys.}
  \textbf{\bibinfo{volume}{B593}}, \bibinfo{pages}{183} (\bibinfo{year}{2001}),
  \eprint{hep-lat/0007004}.

\bibitem[{\citenamefont{Jenkins and
  Manohar}(1991{\natexlab{a}})}]{Jenkins:1991jv}
\bibinfo{author}{\bibfnamefont{E.}~\bibnamefont{Jenkins}} \bibnamefont{and}
  \bibinfo{author}{\bibfnamefont{A.~V.} \bibnamefont{Manohar}},
  \bibinfo{journal}{Phys. Lett.} \textbf{\bibinfo{volume}{B255}},
  \bibinfo{pages}{558} (\bibinfo{year}{1991}{\natexlab{a}}).

\bibitem[{\citenamefont{Jenkins and
  Manohar}(1991{\natexlab{b}})}]{Jenkins:1991es}
\bibinfo{author}{\bibfnamefont{E.}~\bibnamefont{Jenkins}} \bibnamefont{and}
  \bibinfo{author}{\bibfnamefont{A.~V.} \bibnamefont{Manohar}},
  \bibinfo{journal}{Phys. Lett.} \textbf{\bibinfo{volume}{B259}},
  \bibinfo{pages}{353} (\bibinfo{year}{1991}{\natexlab{b}}).

\bibitem[{\citenamefont{Jenkins and
  Manohar}(1991{\natexlab{c}})}]{Jenkins:1991ne}
\bibinfo{author}{\bibfnamefont{E.}~\bibnamefont{Jenkins}} \bibnamefont{and}
  \bibinfo{author}{\bibfnamefont{A.~V.} \bibnamefont{Manohar}}
  (\bibinfo{year}{1991}{\natexlab{c}}), \bibinfo{note}{talk presented at the
  Workshop on Effective Field Theories of the Standard Model, Dobogoko,
  Hungary, Aug 1991}.

\bibitem[{\citenamefont{Gasser and Leutwyler}(1985)}]{Gasser:1985gg}
\bibinfo{author}{\bibfnamefont{J.}~\bibnamefont{Gasser}} \bibnamefont{and}
  \bibinfo{author}{\bibfnamefont{H.}~\bibnamefont{Leutwyler}},
  \bibinfo{journal}{Nucl. Phys.} \textbf{\bibinfo{volume}{B250}},
  \bibinfo{pages}{465} (\bibinfo{year}{1985}).

\bibitem[{\citenamefont{Nozawa and Leinweber}(1990)}]{Nozawa:1990gt}
\bibinfo{author}{\bibfnamefont{S.}~\bibnamefont{Nozawa}} \bibnamefont{and}
  \bibinfo{author}{\bibfnamefont{D.~B.} \bibnamefont{Leinweber}},
  \bibinfo{journal}{Phys. Rev.} \textbf{\bibinfo{volume}{D42}},
  \bibinfo{pages}{3567} (\bibinfo{year}{1990}).

\bibitem[{\citenamefont{Jones and Scadron}(1973)}]{Jones:1973ky}
\bibinfo{author}{\bibfnamefont{H.~F.} \bibnamefont{Jones}} \bibnamefont{and}
  \bibinfo{author}{\bibfnamefont{M.~D.} \bibnamefont{Scadron}},
  \bibinfo{journal}{Ann. Phys.} \textbf{\bibinfo{volume}{81}},
  \bibinfo{pages}{1} (\bibinfo{year}{1973}).

\bibitem[{\citenamefont{Beane and Savage}(2002)}]{Beane:2002vq}
\bibinfo{author}{\bibfnamefont{S.~R.} \bibnamefont{Beane}} \bibnamefont{and}
  \bibinfo{author}{\bibfnamefont{M.~J.} \bibnamefont{Savage}},
  \bibinfo{journal}{Nucl. Phys.} \textbf{\bibinfo{volume}{A709}},
  \bibinfo{pages}{319} (\bibinfo{year}{2002}), \eprint{hep-lat/0203003}.

\end{thebibliography}
\end{document}